\documentclass[twocolumn,aps,prb,longbibliography,superscriptaddress,floatfix]{revtex4-2}

\usepackage[T1]{fontenc}
\usepackage[utf8]{inputenc}
\usepackage[english]{babel}
\usepackage{color}
\usepackage{float}
\usepackage{braket}
\usepackage{amsmath}
\usepackage{amstext}
\usepackage{amssymb}
\usepackage{graphicx}
\usepackage{bbold}
\usepackage[unicode=true,pdfusetitle,
bookmarks=true,bookmarksnumbered=false,bookmarksopen=false,
breaklinks=false,pdfborder={0 0 1},backref=false,colorlinks=false]
{hyperref}

\hypersetup{
	colorlinks,linkcolor=blue,citecolor=blue,urlcolor=blue}
\makeatletter
\date{} 

\makeatother
\definecolor{Blue}{rgb}{0,0.0,1}
\usepackage{babel}

\begin{document} 

\author{Tarik P. Cysne}
\email{tarik.cysne@gmail.com}
\affiliation{Instituto de F\'\i sica, Universidade Federal Fluminense, 24210-346 Niter\'oi RJ, Brazil}

\author{Marcio Costa}
\affiliation{Instituto de F\'\i sica, Universidade Federal Fluminense, 24210-346 Niter\'oi RJ, Brazil} 

\author {Marco Buongiorno Nardelli}
\address{Department of Physics and Department of Chemistry, University of North Texas, Denton TX, USA }

\author{R. B. Muniz}
\affiliation{Instituto de F\'\i sica, Universidade Federal Fluminense, 24210-346 Niter\'oi RJ, Brazil} 
\email{bechara@if.uff.br}

\author{Tatiana G. Rappoport}
\affiliation{Centro de Física das Universidade do Minho e do Porto (CF-UM-UP) e Departamento de Física, Universidade do Minho, P-4710-057 Braga, Portugal}	
\affiliation{Instituto de F\'\i sica, Universidade Federal do Rio de Janeiro, C.P. 68528, 21941-972 Rio de Janeiro RJ, Brazil}

\title{Ultrathin films of black phosphorus as suitable platforms for unambiguous observation of the orbital Hall effect}
 
\begin{abstract}
Phosphorene, a monolayer of black phosphorus, is a two-dimensional material that lacks a multivalley structure in the Brillouin zone and has negligible spin-orbit coupling. This makes it a promising candidate for investigating the orbital Hall effect independently of the valley or spin Hall effects. To model phosphorene, we utilized a DFT-derived tight-binding Hamiltonian, which is constructed with the pseudo atomic orbital projection method. For that purpose, we use the \textsc{paoflow} code with a newly implemented internal basis that provides a fairly good description of  the phosphorene conduction bands. By employing linear response theory, we show that phosphorene exhibits a sizable orbital Hall effect with strong anisotropy in the orbital Hall conductivity for the out-of-plane orbital angular momentum component. The magnitude and sign of the conductivity depend upon the in-plane direction of the applied electric field. These distinctive features enable the observation of the orbital Hall effect in this material unambiguously. The effects of strain and of a perpendicularly applied electric field on the phosphorene orbital-Hall response are also explored. We show that a supplementary electric field applied perpendicular to the phosphorene layer in its conductive regime gives rise to an induced in-plane orbital magnetization.
\end{abstract}
\maketitle 

%%%%%%%%%%%%%%%%%%%%%%%%%%%%%%%%%%%%%%%%%%%%%%%%%%%
%%%%%%%%%------------------------------Sec 1-----------------------------------%%%%%%%%%%%%%%
%%%%%%%%%%%%%%%%%%%%%%%%%%%%%%%%%%%%%%%%%%%%%%%%%%%

\section{Introduction} 
The phenomenon known as the orbital Hall effect (OHE) is characterized by the emergence of an orbital angular momentum (OAM) current that flows transversely to the direction of an applied electric field. Distinctly from the spin Hall effect (SHE), the OHE does not require the presence of spin-orbit interaction to occur. Despite being predicted nearly two decades ago \cite{Bernevig-Hughes-Zhang-PhysRevLett.95.066601}, the prospect of using the OHE to generate OAM current in certain materials has recently sparked great interest in the solid-state physics community \cite{Mele-PhysRevLett.123.236403, Oppeneer-PhysRevMaterials.6.095001, Salemi-PhysRevB.106.024410, OH-Torque-PhysRevB.107.134423, go2023intrinsic, Mokrousov-PhysRevMaterials.6.074004, HW-Lee-PhysRevLett.128.176601, fonseca2023orbital, Gambardella-PhysRevResearch.4.033037, urazhdin2023symmetry, busch2023orbital,m2023spin}. OAM currents can be produced in a wide range of materials, and their intensities can exceed those of spin current. Furthermore, they can be injected into adjacent elements to exert torque on magnetic units, expanding their possible applications in orbitronics \cite{Go-Hyun-Woo-PhysRevLett.121.086602, Go_EPL-Review}.  

As a matter of fact, light metals with weak spin-orbit coupling are being explored as a means of generating orbital currents in three-dimensional metals \cite{Go-Experiment-https://doi.org/10.48550/arxiv.2109.14847}. Recently, orbital torques have been realized in light metal/ferromagnet heterostructures, providing indirect but robust experimental evidence of the OHE \cite{Zheng-OrbTorque-PhysRevResearch.2.013127, Lee-OrbTorque-10.1038/s42005-021-00737-7,Lee2021}.  

The OHE has also been investigated in two-dimensional (2D) materials that, in some cases, may host an orbital Hall (OH) insulating phase, characterized by a finite OH conductivity plateau located within the insulating band gap \cite{Canonico-PhysRevB.101.161409, Canonico-PhysRevB.101.075429}. Recent studies have shed light on the fascinating properties of the OH insulating phase in these materials, such as its connection with higher order topological phases \cite{Costa2023} and the encoding of non-trivial topology associated with OAM in an orbital Chern number \cite{Cysne-PhysRevLett.126.056601, Cysne-PhysRevB.105.195421}. 

The difficulty in discerning the OHE from other angular-momentum transport phenomena has hindered its unequivocal direct observation. For example, in some cases the spin accumulation produced by the spin Hall effect may be hard to distinguish from its orbital angular momentum counterpart. The valley Hall effect (VHE) induced by a longitudinally applied electric field that occur in non-centrosymmetric lattices with multi-valley structure in the Brillouin zone involves the transverse flow of valley currents that may also carry magnetic moment \cite{Cysne-PhysRevLett.126.056601, Cysne-PhysRevB.105.195421, Bhowal-PhysRevB.103.195309, Salvador-Sanchez-https://doi.org/10.48550/arxiv.2206.04565}, which can be hard to dissociate from the intra-atomic orbital Hall contribution.      

Multi-orbital 2D materials possess natural symmetry constraints that lead to various types of orbital hybridization, which can maximize the OHE~\cite{Costa2023}. However, to single out the OHE unequivocally it is crucial to identify materials with weak spin-orbit coupling that display no significant spin Hall effect (SHE), nor VHE or magnetoelectric effects that may mask the OHE.

In this article, we suggest that phosphorene is a very suitable material for direct observing the OHE in 2D materials. It is a centrosymmetric semiconductor with a sizeable direct band gap at the $\Gamma$ point of the 2D BZ \cite{Symmetry-Phosphorene-PhysRevB.90.115439, Phosphorne-SpectraStrain-PhysRevLett.112.176801, Phosphorene-Spectra-strain-PhysRevB.94.085417,Phosphorene-ModelRudenko-PhysRevB.89.201408, Paulo-Fabian-PhysRevB.100.115203} that does not host VHE. In the absence of reasonably strong electric fields, applied perpendicularly to the layer, it behaves as an ordinary insulator  and  shows no spin Hall effect (SHE) within its band gap \cite{ElectricField-Fazzio-Zunger-doi:10.1021/nl5043769}. The spin-orbit interaction in phosphorene is extremely weak \cite{Popovic2015, Avsar2017} and consequently it also displays negligible SHE in the metallic regime in comparison with the OHE, as we shall see later. Symmetry prevents the appearance of the magneto-electric effect in phosphorene, even in the presence of strain \cite{Hu-SymmetriesPolarization-doi:10.1021/acs.nanolett.6b04630}. 

Here, we have performed density functional theory (DFT) calculations combined with linear-response theory to analyze the OH response in phosphorene. Our calculations show that phosphorene exhibit sizeable anisotropic OH conductivities that change sign for in-plane electric fields applied along the armchair ($\hat{x}$) and zigzag ($\hat{y}$) Cartesian directions depicted in Fig. \ref{fig:fig1}. These features persist in the presence of moderate in-plane deformation and of electric fields applied perpendicular to the monolayer, i.e., along $\hat{z}$. Furthermore, we show that the presence of perpendicular electric fields may lead to the appearance of a current-induced orbital magnetization oriented parallel the phosphorene layer.

\section{DFT derived Hamiltonian \label{Sec:Phosphorene}}  
\begin{figure*}[t]
	\centering
	 \includegraphics[width=0.75\linewidth,clip]{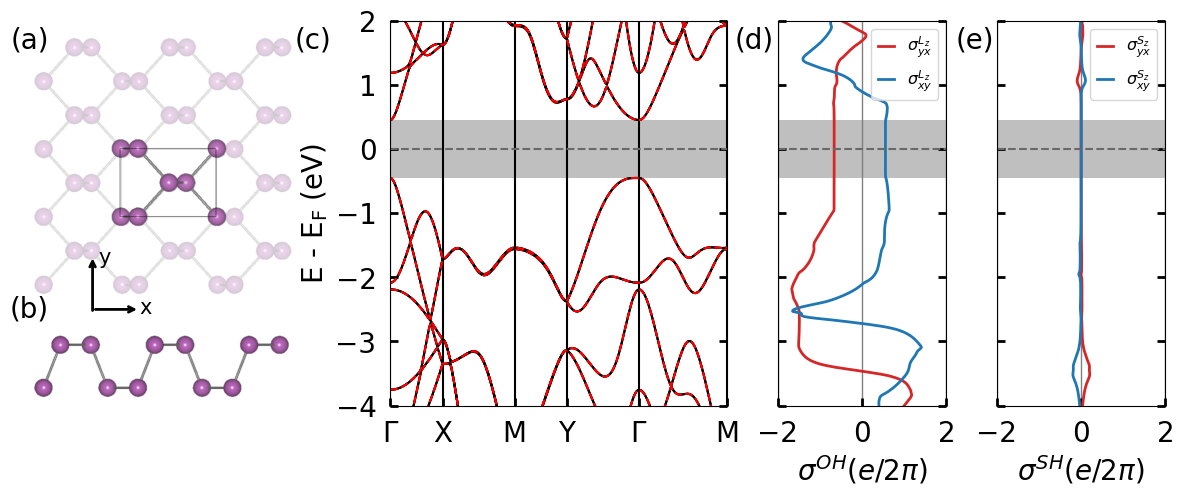}     
	\caption{Phosphorene crystal structure (a) top and (b) side views. Band-structure (c) DFT (solid black line) and \textsc{paoflow} (dashed red line). Orbital Hall (OH) (d) and the spin Hall (SH) (e) conductivities calculated as functions of energy for electric fields applied along the $\hat{x}$ (red) and $\hat{y}$ (blue) directions. }
	\label{fig:fig1}
\end{figure*}

Phosphorene is a two-dimensional material composed of a single layer of phosphorus atoms arranged in a distorted honeycomb lattice structure (figure \ref{fig:fig1}(a)), similar to graphene. However, unlike graphene, the lattice structure of phosphorene is puckered, with a non coplanar configuration as illustrated in Figure \ref{fig:fig1}(b).

Our DFT calculations \cite{DFT1,DFT2} were carried out with the plane-wave-based code \textsc{Quantum Espresso} \cite{QE-2017} to compute the band structure and eigenstates of phosphorene. The generalized gradient approximation (GGA) \cite{PBE} was used to treat the exchange and correlation potential, while fully relativistic projected augmented wave (PAW) potentials \cite{PAW,pslibrary} were employed to describe the ionic cores. To ensure accurate results, we set the wavefunctions cutoff energy to 44 Ryd and the charge density cutoff energy is ten times larger. Our self-consistent calculations (SCF) were executed with a linear density of $\bf k$-points of 12.0/\AA$^{-1}$ in the 2D Brillouin zone, and a minimum of 15 \AA{} of vacuum is taken to avoid spurious interactions. We included a static electrical field (along the z direction) using a full SCF calculation via the modern theory of polarization \cite{Efield}

Figure \ref{fig:fig1}(c), shows the band structure of phosphorene displaying its direct bandgap at the $\Gamma$ point. Phosphorene's puckered crystalline structure is highly anisotropic, as evidenced by its energy spectrum near $\Gamma$, which presents a parabolic dispersion along the $\Gamma$-Y direction and a linear behavior along $\Gamma$-X. Furthermore, the puckering of the lattice has a notable impact on the mechanical and electronic characteristics of phosphorene. It renders the material more susceptible to strain, as deformation can significantly alter its bandgap and electronic transport properties \cite{Phosphorne-SpectraStrain-PhysRevLett.112.176801}. 

To perform linear response calculations, we utilized the pseudo atomic orbital projection method \cite{PAO1,PAO2, PAO3, PAO4} implemented in the \textsc{paoflow} code \cite{PAO5,PAO6}. This approach involves constructing an effective tight-binding Hamiltonian, with no adjustable parameters, from the DFT calculations. In general, we project the plane-wave Kohn-Sham orbitals onto the compact subspace spanned by the pseudo atomic orbitals (PAO), which are naturally included in the PAW potentials. The vast majority of cases can be accurately described by this approach with an excellent agreement between the DFT and \textsc{paoflow} band-structure. Nevertheless, occasionally the PAO basis fails to reproduce the conduction bands, especially when the unoccupied bands have a relatively strong character of an orbital that is not included in the PAO base, as in the case of phosphorene. Its conduction, and to a minor degree the valence bands, are highly hybridized with $d$-orbitals \cite{MENEZES2018411}. Since the pseudo potential used in the calculation (P.rel-pbe-n-kjpaw\_psl.1.0.0.UPF) is generated only with $s$ and $p$ orbitals, this original approach fails. To circumvent this problem, we used the recently implemented \textsc{paoflow} internal basis, which is constructed by solving the atomic DFT problem for an all electron configuration up to desired orbital. Once the atomic wavefunction is obtained the DFT plane-wave wavefunctions are projected as described in ref. \cite{PAO4}.

Figure \ref{fig:fig1}(c) shows the effective tight-binding and the DFT band-structure calculations superimposed. This approach significantly reduces the computational cost of performing large $\bf k$-space numerical integration. We have previously used this method to investigate distinct characteristics of different systems, such as: spin dynamics~\cite{adatoms,fegete}, as well as transport~\cite{hoti, cri3-graphene} and topological properties~\cite{Costa2019, Costa2018}. The orbital Hall conductivity calculations were performed with a reciprocal space sampling that is ten times larger than the one used in our DFT-SCF calculations.

\section{OHE calculations \label{Sec:OHECalculation}}  
Within linear response theory, the current density of angular momentum with polarization $\eta$, flowing along the $\mu$ direction ($\mathcal{J}^{X_{\eta}}_{\mu} $), can be generically expressed in terms of the angular momentum conductivity tensor by $\mathcal{ J}^{X_ { \eta}}_{\mu}=\sum_{\nu}\sigma^{X_{\eta}}_{\mu,\nu} \mathcal{E}_{\nu}$. Here, $\mathcal{E}_{\nu}$ symbolizes the $\nu$-component  of the applied electric field; $\eta$, $\mu$ and $\nu$ label  the Cartesian components $x,y,z$. $X_{\eta}$ represents the $\eta$-component of either the orbital angular momentum operator ($\hat{\ell}_{\eta}$) or the spin  operator ($\hat{s}_{\eta}$), depending on the nature of the induced angular momentum that drifts. The conductivity tensor is given by
\begin{eqnarray}
\sigma^{X_{\eta}}_{\mu,\nu}=\frac{e}{(2\pi)^2}\sum_{n} \int_{BZ} d^2{\bf k} f_{n {\bf k}} \Omega_{\mu,\nu , n}^{X_{\eta}} ({\bf k}),
\label{conductivity}
\end{eqnarray}
where, the orbital (spin) Berry curvature 
\begin{eqnarray}
{\Omega_{\mu,\nu , n}^{X_{\eta}} ({\bf k})}= {2\hbar}\sum_{m\neq n}\text{Im} \Bigg[ \frac{\langle u_{n,{\bf k}}\big|j_{\mu,{\bf k}}^{X_{\eta}}\big|u_{m,{\bf k}}\rangle \langle u_{m,{\bf k}}\big|v_{\nu,{\bf k}}\big|u_{n,{\bf k}}\rangle}{(E_{n,{\bf k}}-E_{m,{\bf k}}+i0^+)^2}\Bigg]. \nonumber \\
\label{OBerry}
\end{eqnarray}
The $\nu$-component of the velocity operator may be obtained by $v_{\nu,\bf k}=\hbar^{-1}\partial \mathcal{H} ({\bf k})/\partial k_{\nu}$, where $\mathcal{H} ({\bf k})$ represents the Hamiltonian in reciprocal space, and ${\bf k}$ stand for the wave vector. Here, $\big|u_{n,{\bf k}}\rangle$ is the periodic part of the Bloch eigenstate of $\mathcal{H} ({\bf k})$, associated with band energy $E_{n, {\bf k}}$ and $f_{n {\bf k}}$ symbolizes the Fermi-Dirac distribution function. The orbital (spin) angular momentum current operator that flows along the $\mu$-direction with orbital (spin) polarization in the $\eta$-direction, is defined by $j_{\mu,{\bf k}}^{X_{\eta}}=\left(X_{\eta}v_{\mu, {\bf k}}+v_{\mu, {\bf k}}X_{\eta} \right)/2$, where $X_{\eta}=\hat{\ell}_{\eta} (\hat{s}_{\eta})$. 

Our calculations are performed at zero temperature, but they are expected to be valid at room temperature because the energy band gap of phosphorene is relatively high ($\approx$ 1eV). 
 
\section{Results and Discussion} 
%\subsection{Phosphorene}
Figure \ref{fig:fig1}(d) shows the orbital Hall conductivities $\sigma^{L_z}_{xy}$ and $\sigma^{L_z}_{yx}$, calculated as functions of Fermi energy, for in-plane electric fields applied along the $\hat{y}$ and $\hat{x}$ directions, respectively. Both conductivities present a plateaux inside the energy-band gap. These orbital Hall conductivity plateaux are robust against disorder due to the lack of a Fermi surface in this energy region \cite{liu2023dominance, DimitrovaVertex-PhysRevB.71.245327}. Phosphorene has been proposed to be a higher-order topological insulator \cite{HOTI-Phosphorene-PhysRevB.104.125302, HOTI-Phosphorene-PhysRevB.98.045125}, a type of topological state that was recently connected to the orbital Hall insulating phase \cite{Costa2023}. We note that $\sigma^{L_z}_{xy}$ is markedly different from $\sigma^{L_z}_{yx}$ inside and close to the energy-band gap, where they have opposite signs. This reflects the high anisotropy of the phosphorene lattice structure. The rectangular unit cell of phosphorene illustrated in Fig. \ref{fig:fig1}(a) indicates that $\hat{x}$ and $\hat{y}$ directions are not equivalent. This is confirmed by Figure \ref{fig:fig1}(c), which shows that the valence band dispersion relations along the $\Gamma$-X and $\Gamma$-Y directions are markedly different. Hence, the carrier group velocities, their effective masses, as well as the orbital current densities are expected to be different for electric fields applied along the $\hat{x}$ and $\hat{y}$ directions. The crystalline symmetry of phosphorene also ensures that in-plane electric fields can only induce transverse currents of angular momentum polarized along $\hat{z}$. This holds for both orbital and spin angular momentum currents, because they are subjected to essentially the same crystalline symmetry constraints \cite{H-Woo-Symmetry_PhysRevB.105.035142}. In a crystal with a given space group, the spin and orbital Berry curvatures must be invariant under all symmetry operations of the group. This means that if a given symmetry operation, such as rotation, mirror reflection, or spatial inversion, changes the sign of the spin or orbital Berry curvature, then the corresponding component of the spin or orbital Hall conductivity is forbidden by symmetry. The presence or absence of symmetries in the crystal structure can dictate which components of the Hall conductivity are allowed or forbidden (see Appendix A). 

\begin{figure}[h!]
	\centering	 \includegraphics[width=0.75\linewidth,clip]{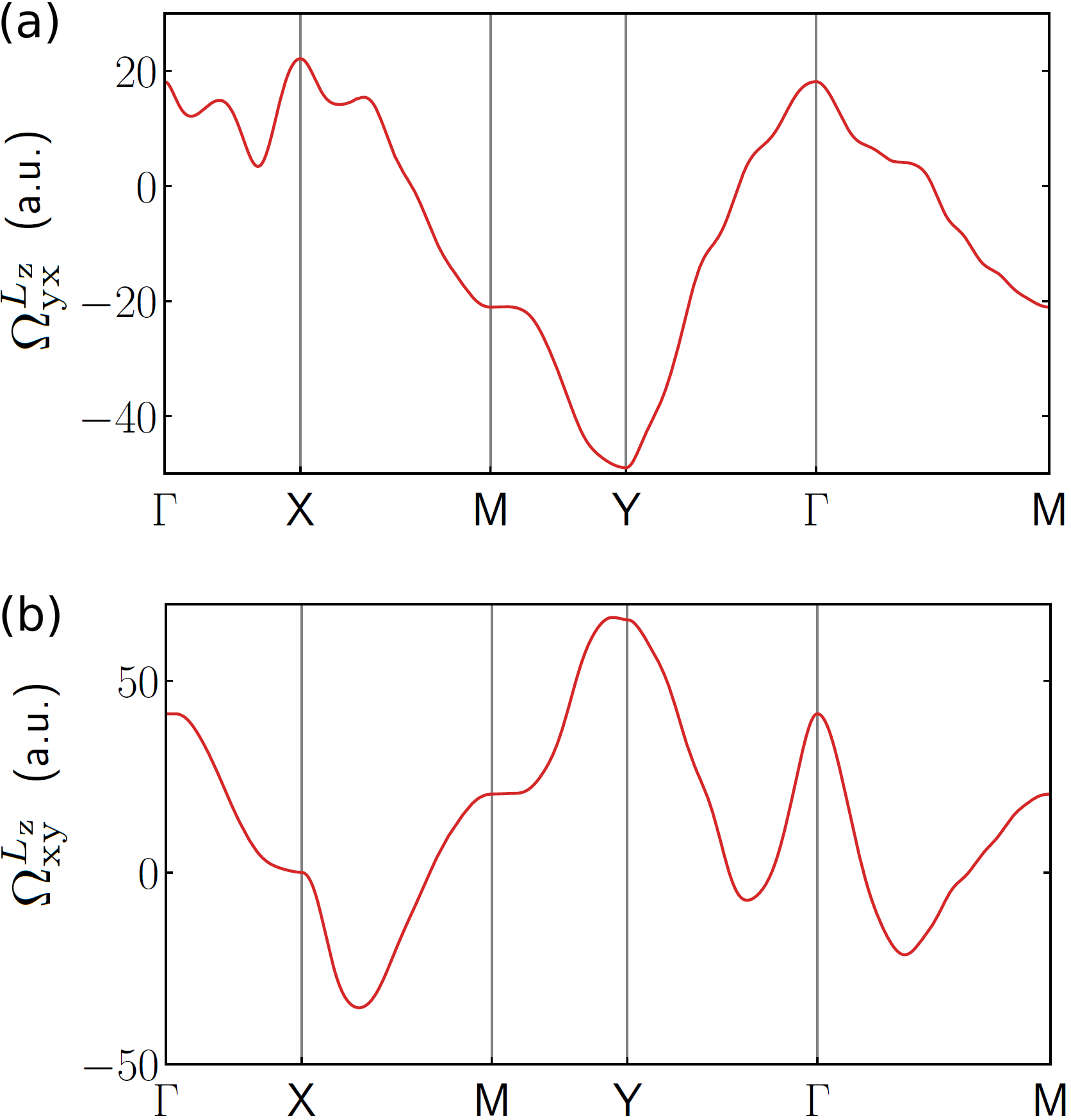}   
	\caption{Orbital Berry curvatures summed over occupied bands of phosphorene $\Omega^{L_z}_{yx}({\bf k})$ (a) and $\Omega^{L_z}_{xy}({\bf k})$ (b) calculated along high symmetry directions of the Brillouin zone.}
	\label{fig:figOBC}
\end{figure}

It is instructive to inquire into the regions of the 2D Brillouin zone (BZ) that contribute most to $\sigma^{L_z}_{xy}$ and $\sigma^{L_z}_{yx}$. For this purpose we have calculated the orbital-Berry curvatures summed over the occupied bands of phosphorene, computed as functions of the wave-vector $\bf{k}$ along some high-symmetry directions of the 2D BZ. The results are depicted in Fig. \ref{fig:figOBC}, and they clearly show that the highest contributions to both $\sigma^{L_z}_{xy}$ and $\sigma^{L_z}_{yx}$ come from the region around the Y-symmetry point. Correlating the orbital Berry curvature with the band structure and the onset of the orbital Hall conductivity at energies below the gap, one can notice that the main contribution to the orbital Hall plateau originates from the transition from lower bands to the higher valence band in the region around the Y-symmetry point, at energies well below the top of the valence band, which is located in $\Gamma$.

The change of sign in the phosphorene OH-conductivity may be experimentally verified by observing the induced orbital magnetic moment accumulations on the boundaries of phosphorene samples, similar to SHE experiments \cite{SHE-Devices-Jungwirth2012, Go-Experiment-https://doi.org/10.48550/arxiv.2109.14847, marui2023spin,kumar2023ultrafast,lyalin2023magnetooptical}. The small spin-orbit coupling and the topological triviality of phosphorene, with respect to $\mathbb{Z}_2$, make the SHE orders of magnitude smaller than the OHE [see Fig. \ref{fig:fig1} (e)]. In addition, the electronic spectrum of phosphorene has no multivalley structure in the 2D Brillouin zone and hence does not host VHE. Thus, phosphorene offers an ideal platform for unambiguous observation of the OHE. We note that the OH conductivity of phosphorene is of the same order of magnitude as those predicted for monolayers of transition metal dichalcogenides \cite{Canonico-PhysRevB.101.161409, Costa2023}. 
It is noteworthy that the OHE increases with the number of layers \cite{Cysne-PhysRevLett.126.056601, Cysne-PhysRevB.105.195421, Cysne-PhysRevB.107.115402} and so, thin films of black phosphorus may be employed to enhance the OH signal in such experiments. However, one must keep in mind that the band gap decreases monotonically with the increase in the number of layers, saturating at approximately 0.3 eV for sufficiently large film thicknesses \cite{ElectricField-Fazzio-Zunger-doi:10.1021/nl5043769}.

In general, the transport properties of 2D materials are influenced by the substrate, which may cause strain and/or alter the features of the sample's surface in contact with it. In some cases it is necessary to encapsulate the film to prevent its deterioration from oxidation and also be able to control its density of carriers with gate voltages. Therefore, it is worth investigating how strain and the presence of an auxiliary perpendicular electric field would affect the orbital transport properties of phosphorene.  

\subsection{Effects of Strain}

\begin{figure}[h!]
	\centering	 \includegraphics[width=1\linewidth,clip]{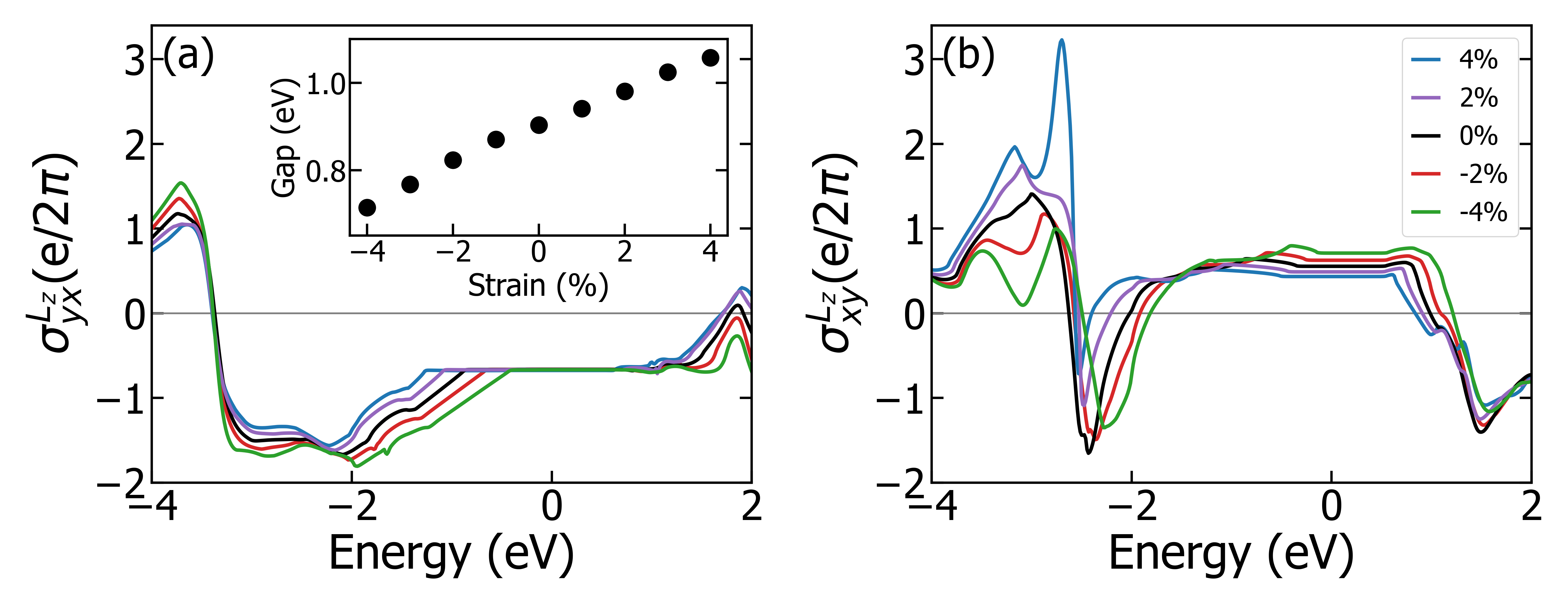}   
	\caption{Orbital Hall conductivity $\sigma_{yx}^{L_z}$ (a) and $\sigma_{xy}^{L_z}$ (b), calculated as functions of energy, for different strains (compressive and tensile) along the x-axis. The inset in panel (a) shows the strain-induced change in the band-gap size.}
	\label{fig:fig2}
\end{figure} 

Figure \ref{fig:fig2} illustrates the effects of uniaxial strain (both compressive and tensile) along the $\hat{x}$ direction, on the OH conductivity components $\sigma^{L_z}_{xy}$ and $\sigma^{L_z}_{yx}$. 

When subjected to moderate in-plane uniform strain, the unit cell of phosphorene maintains its rectangular shape, as long as it does not undergo structural phase transitions, which typically require high levels of strain \cite{GroupTheory-PhysRevB.91.205421, Structural-Phase-PhysRevB.92.064114}. Thus, with the values of strain used in Fig. \ref{fig:fig2}, the $D_{\text{2h}}$ group symmetry of phosphorene is preserved and hence only the $L_z$ component of the OHC remains non-null. Strain clearly affects the OH conductivity of phosphorene. It modifies the electronic states around the band gap \cite{Peng-Wei-Copple-PhysRevB.90.085402}, may alter their orbital features and the orbital transport in general. 
It is noteworthy that the plateau heights of $\sigma_{xy}^{L_z}$ and $\sigma_{yx}^{L_z}$ do not change significantly under moderate deformations along the x-direction, showing that the OHE is reasonably robust to uniaxial strain within the energy band gap of phosphorene. On the other hand, the length of the OHC plateaux decrease (increase) under compressive (tensile) strain, which is expected because the energy band-gap size follows the same trend \cite{Lewenkopf-Midtvedt2016}, as illustrated in the inset of Fig. \ref{fig:fig2}(a).

\subsection{Effect of Perpendicular Electric-Field}

\begin{figure}[h!]
	\centering	 \includegraphics[width=1\linewidth,clip]{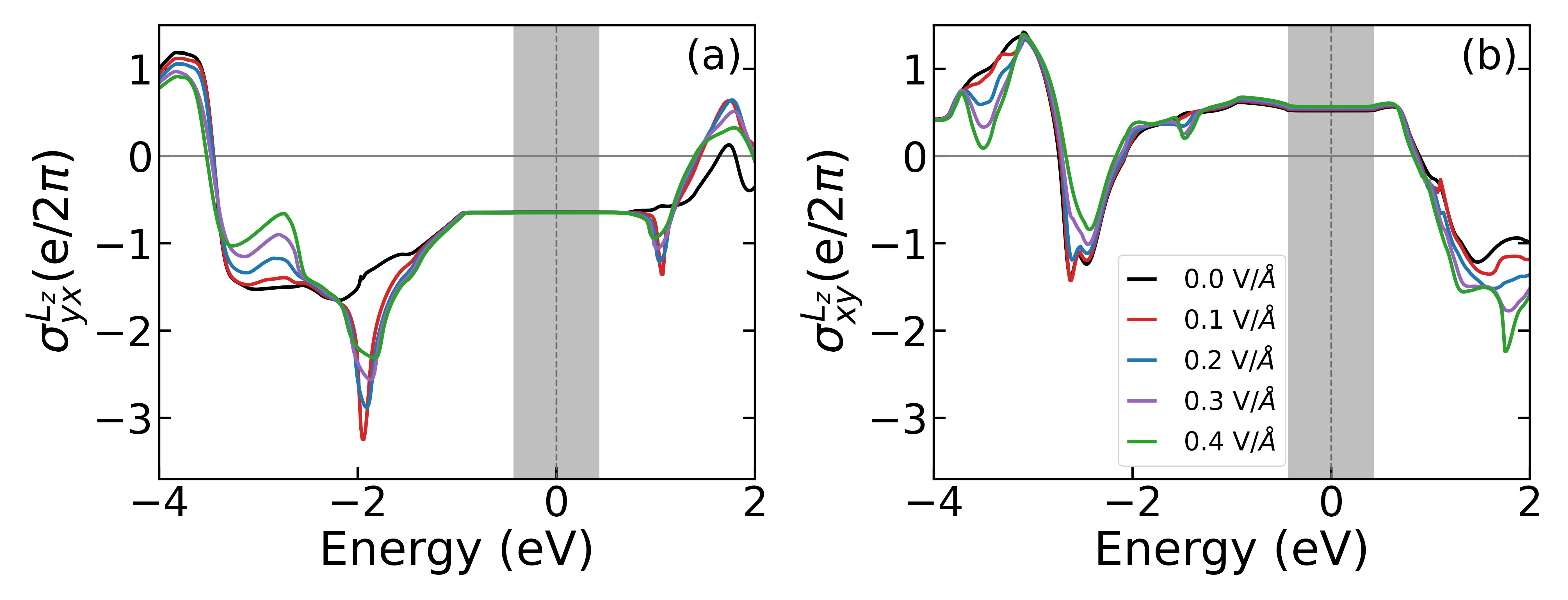}   
	\caption{Orbital Hall conductivity $\sigma_{yx}^{L_z}$ (a) and $\sigma_{xy}^{L_z}$ (b), calculated as functions of energy, for different values of the perpendicularly applied electric field $E_\perp$. } 
	\label{fig:fig3}
\end{figure} 

\subsubsection{Orbital Hall Conductivity}
We shall now examine how the OHC of phosphorene is affected by an electric field $\vec{E}_\perp = E_\perp \hat{z}$, applied perpendicularly to its layer. The presence of $\vec{E}_\perp$ reduces the phosphorene point group $D_{{\rm 2h}}$ to $C_{{\rm 2v}}$, which belong to the same Laue class $mmm$. Since the Laue class determines the general form of the OHC tensor \cite{Ebert-Symmetry-tensor, Marcosguimaraes}, only the $L_{z}$ component of the OHC remains non-null in the presence of the $\vec{E}_\perp$ (see Appendix A). 

Figure \ref{fig:fig3} shows $\sigma^{L_z}_{yx}$ and $\sigma^{L_z}_{xy}$, calculated as functions of energy, for different values of $E_\perp$. We note that the OHC is much more affected by $E_\perp$ in some energy ranges outside the band gap than within it. We recall that ultrathin films of black phosphorus can switch to a topological insulating phase for sufficiently high values of $E_\perp$, as  discussed in the \cite{ElectricField-Fazzio-Zunger-doi:10.1021/nl5043769}. However,  for phosphorene, this phase transition requires values of $E_\perp \gg 0.6$V/m, which is higher than the ones considered in Fig. \ref{fig:fig3}.  

\subsubsection{Orbital Magnetoelectric Effect} 

The noncentrosymmetric and polar $C_{\rm 2v}$ point group allows the occurrence of orbital magnetoelectric effect mediated by Fermi-surface conducting states \cite{C2v-Magnetoelectric-PhysRevResearch.3.023111, Cysne-PhysRevB.107.115402, Cysne-OMEpxpy-PhysRevB.104.165403,Koki-Peters-PhysRevB.107.214109, Koki-Peters-PhysRevB.107.094106, Hayami-PhysRevB.98.165110}. The perpendicular electric field $\vec{E}_\perp$ distorts the phosphorene's charge distribution, giving rise to a finite polarization $\vec{P}=P_z\hat{z}$ perpendicular to its layer \cite{Hu-SymmetriesPolarization-doi:10.1021/acs.nanolett.6b04630}. The driving field in the phosphore plane exerts a torque on the electric dipoles, thereby inducing a net orbital magnetization $\vec{M}^L\propto \vec{P}\times \vec{\mathcal{E}}$ \cite{Cysne-PhysRevB.107.115402, Cysne-OMEpxpy-PhysRevB.104.165403, Hayami-JPCM-2016, Salemi-Oppeneer-2019}. One may calculate $\vec{M}^L$ utilizing a scheme similar to the one described in the Secs. \ref{Sec:Phosphorene} and \ref{Sec:OHECalculation}. Since time-reversal symmetry is preserved, there are no interband contributions to the orbital magnetoelectric effect in phosphorene. Thus, to first order in the in-plane driving field and for finite values of $E_\perp$, the current-induced orbital magnetization per unity cell area of phosphorene is given by $m_{L_\eta}=\sum_{\nu} \alpha_{\eta\nu} \mathcal{E_{\nu}}$ \cite{Yoda2018-OME, He2020-OME}, where
\begin{eqnarray}
\alpha_{\eta\nu}=&&\frac{e\mu_B}{2\Gamma}\sum_n\int_{\rm BZ}\frac{d^2{\bf k}}{(2\pi)^2} \frac{\partial f_{n,{\bf k}}}{\partial E} \nonumber \\
&& \ \ \times \bra{u_{n,{\bf k}}} v_{\nu,{\bf k}} \ket{u_{n,{\bf k}}}\bra{u_{n,{\bf k}}} \hat{\ell}_{\eta} \ket{u_{n,{\bf k}}}
\end{eqnarray}
represent the matrix elements of the magnetoelectric tensor. Here, $\mu_B$ is the Bohr magneton and $\Gamma$ is the energy scale associated with the electronic relaxation time $\tau_e=\hbar/2\Gamma$. This phenomenological parameter simulates  effects of scattering by inhomogeneities and thermal effects due to phonons. In our calculations we have used $\Gamma=1.6$ meV, that correspond to $\tau_e \approx 200$ fs \cite{Avsar2017}. 

\begin{figure}[h!]
	\centering	 \includegraphics[width=1\linewidth,clip]{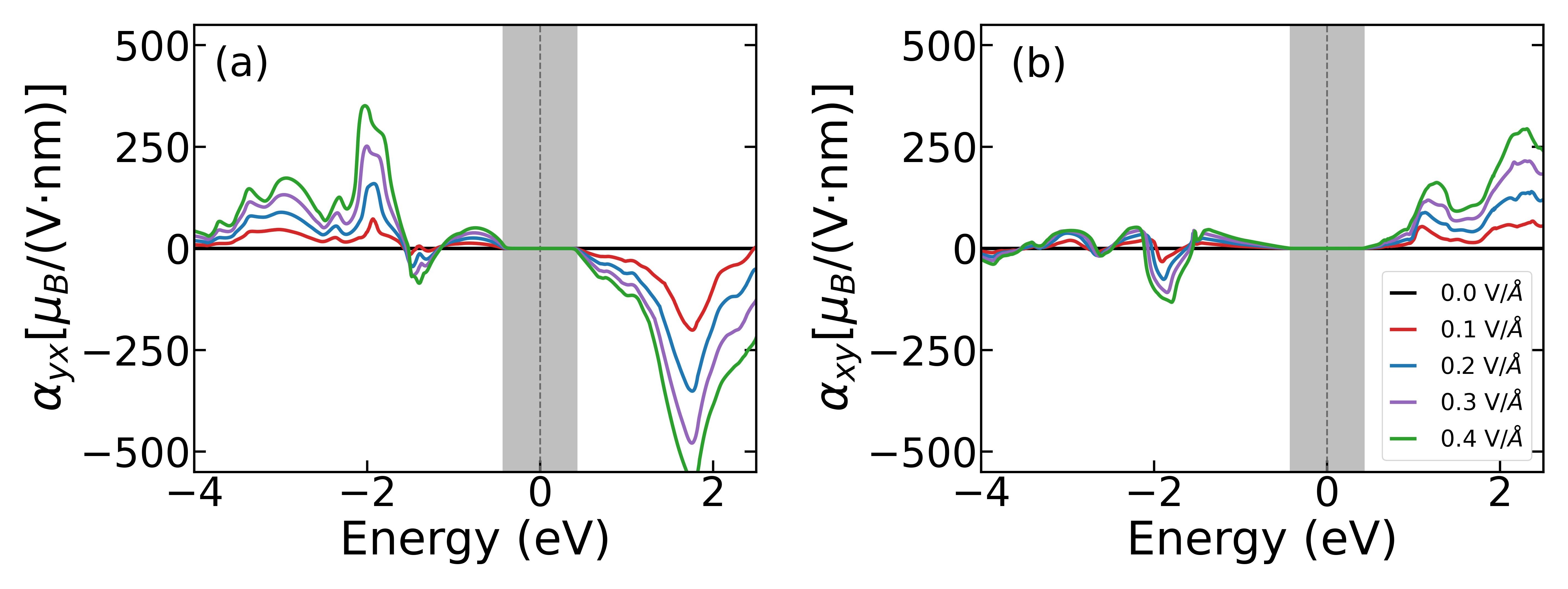}   
	\caption{Orbital magnetoelectric coefficients $\alpha_{yx}$ (a) and $\alpha_{xy}$ (b), calculated as functions of energy, for different values of the perpendicularly applied electric field $E_\perp$.}
	\label{fig:fig4}
\end{figure} 
Fig. \ref{fig:fig4} shows $\alpha_{xy}$ and $\alpha_{yx}$ calculated as functions of energy for different values of the $E_\perp$. As expected, the OME clearly vanishes within the band-gap energy range. However, in the conductive regime, it can reach sizeable values for both $m_{L_y}$ and $m_{L_x}$, in response to electric fields applied along the $\hat{x}$ and $\hat{ y }$ directions, respectively. This inplane-induced orbital magnetization adds up to the orbital angular-momenta accumulated at the sample's edges, due to the OHE, transforming its original antiferromagnetic-like disposition into a non-collinear orbital magnetic arrangement. 

In some energy ranges the OME varies appreciably with $E_\perp$, which may be used to control the OME intensity. In order to roughly estimate the order of magnitude of the in-plane OME we consider an electric field with intensity $\mathcal{E}_x=10^5$ V/m and a carrier density that leads to $\alpha_{yx}=-2\times 10^2 \mu_B/(\text{V.nm})$. In this case, the induced orbital magnetization $m_{L_y} \approx -0.3 \times 10^{-2}\mu_B/A_{\rm u.c.}$, where $A_{\rm u.c.}=0.152 \text{nm}^2$ represents the phosphorene unity cell area. This has the same order of magnitude of the Edelstein effect estimated for Bi/Ag(111) in Ref. \cite{IngridMerting-PhysRevB.97.085417} assuming a larger value of $\tau_e$.

%%%%%%%%%%%%%%%%%%%%%%%%%%%%%%%%%%%%%%%%%%%%%%%%%%%%%
%%%%%%%%%------------------------------Sec 3------------------------%%%%%%%%%%%%%%
%%%%%%%%%%%%%%%%%%%%%%%%%%%%%%%%%%%%%%%%%%%%%%%%%%%%%

\section{Final Remarks and Conclusions} 

To summarize, we argue that thin films of black phosphorus may provide suitable 2D platforms for direct observation of the orbital Hall effect. To this end, we combine linear response theory with density functional theory calculations to investigate the orbital conductivity of phosphorene and explore how it is affected by uniform strain and perpendicular electric fields. We show that phosphorene displays a fairly large OHC, with perpendicular orbital polarization, which is orders of magnitude larger than the SHC. This OHC is also highly anisotropic with respect to the direction of the in-plane applied electric field, and may even switch sign when the driving field direction is changed. Inside the energy band gap, it exhibits an orbital Hall insulating plateau that is robust under moderate uniform strain and perpendicular electric fields. The latter breaks \textcolor{blue}{spatial} inversion symmetry and may lead to the appearance of an in-plane orbital-magnetization, induced by an in-plane electric current. This effect alters the anti-symmetric profile of the orbital magnetic moment induced by the orbital Hall effect in the conducting phase. Our numerical calculations are complemented by symmetry analysis.

\begin{acknowledgments}
	We acknowledge CNPq/Brazil, CAPES/Brazil, FAPERJ/Brazil,  INCT Nanocarbono and INCT Materials Informatics for financial support. TGR acknowledges funding from FCT-Portugal through Grant No. CEECIND/07471/2022. She thankfully acknowledges the computer resources at MareNostrum and the technical support provided by Barcelona Supercomputing Center (FI-2020-2-0033). MC acknowledges CNPq (Grant No. 317320/2021-1) and FAPERJ/Brazil (Grant No. E26/200.240/2023). We thank Profs. A. Fazzio and P. Venezuela for fruitful discussions. We also thank Prof. Davide Ceresoli for implementing the PAOFLOW internal basis.
\end{acknowledgments}	

%%%%%%%%%%%%%%%%%%%%%%%%%%%%%%%%%%%%%%%%%%%%%%%%%%%%%
%%%%%%%%%------------------------------Appdx-----------------------------------%%%%%%%%%%%%%%
%%%%%%%%%%%%%%%%%%%%%%%%%%%%%%%%%%%%%%%%%%%%%%%%%%%%%

\appendix
\section{Symmetry constraint on orbital Hall conductivity \label{AppSymm}}

The crystal symmetry operations of phosphorene are: $E$, $\tau\mathcal{C}_{2x}$, $\mathcal{C}_{2y}$, $\tau\mathcal{C}_{2z}$, $\mathcal{P}$, $\tau \mathcal{M}_x$, $\mathcal{M}_y$ and $\tau \mathcal{M}_z$ \cite{Symmetry-Phosphorene-PhysRevB.90.115439}. Here $E$ represents the identity operation, $\mathcal{C}_{2\mu}$ is a 180$^o$ rotation around the $\mu$-axis, $\mathcal{M}_{\mu}$ denotes a reflection through a mirror plane that is perpendicular to the $\mu$-axis, and $\mathcal{P}$ symbolizes the spatial-inversion operation; $\tau\mathcal{O}$ designates the action of $\mathcal{O}$ followed by a half-unity-cell translation $\vec{\tau}=(a_x/2,a_y/2)$, where $a_x$ and $a_y$ represent the moduli of the unit cell lattice vectors. This set of symmetries is isomorphic to the point group $D_{{\rm 2h}}$, which correspond to Laue group \emph{mmm} \cite{Ebert-Symmetry-tensor}. Consequently, for a phosphorene layer in the $xy$ plane, only the $L_z$ component of the OHE is allowed \cite{Marcosguimaraes}. It is possible to derive the constraints to the OH conductivity tensor imposed by each symmetry operation of phosphorene. They are summarized in the table \ref{tableSym}. 

\begin{table}[h!]
	\centering
	\begin{tabular}{||c ||c c | c c | c c ||} 
		\hline
		Symmetry & & $\sigma^{L_x}_{{\rm OH}}$ & &  $\sigma^{L_y}_{{\rm OH}}$ & &  $\sigma^{L_z}_{{\rm OH}}$ \\ [0.5ex] 
		\hline
		\hline
		$\mathcal{P}$ & & $\bigcirc$ & & $\bigcirc$ & & $\bigcirc$ \\ \hline
		$\tau \mathcal{M}_z$,  $\tau \mathcal{C}_{2z}$$^{(*)}$ & & $\times$ & & $\times$ & & $\bigcirc$ \\ \hline
		$\tau \mathcal{M}_x$$^{(*)}$, $\tau \mathcal{C}_{2x}$ & & $\times$ & & $\bigcirc$ & & $\bigcirc$ \\ \hline
		$\mathcal{M}_y$$^{(*)}$, $\mathcal{C}_{2y}$ & & $\bigcirc$ & & $\times$ & & $\bigcirc$ \\ [0.5ex] \hline
	\end{tabular}
	\vskip 2.5 pt
	\caption{Constraints on the OHC of phosphorene imposed by crystal symmetry. We assume that the phosphorene layer lies on the xy-plane, with the $\hat{x}$ and $\hat{y}$ axes oriented as illustrated in Fig. \ref{fig:fig1}. The constraints hold for both $\sigma^{L_{\eta}}_{xy}$ and $\sigma^{L_{\eta}}_{yx}$, which are generically represented here by $\sigma^{L_{\eta}}_{{\rm OH}}$.  
 The OHC components that are allowed (forbiden) by symmetry are identified by $\bigcirc$ ($\times$). 
 The crystal symmetry operations of phosphorene in the presence of $\vec{E}_\perp$ are identified with an asterisk.}
	\label{tableSym}
\end{table} 

In the presence of $\vec{E}_\perp$ all symmetry operations that interchange $z$ and $-z$ are excluded, leaving just $\tau\mathcal{C}_{2z}$, $\tau \mathcal{M}_x$, $\mathcal{M}_y$, and $E$, which are identified with an asterisk in table \ref{tableSym}. In this case, the point group is reduced from $D_{\rm 2h}$ to $C_{\rm 2v}$. However, since $C_{\rm 2v}$ and $D_{\rm 2h}$  belong to the same Laue class ($mmm$), only the $L_z$ component of the OHC can be non zero when phosphorene is subjected to $\vec{E}_\perp$\cite{Ebert-Symmetry-tensor, Marcosguimaraes}.

In order to obtain the constraints on the  OHC components presented in Table \ref{tableSym} we consider the action of $\tau\mathcal{O}$ on the Bloch eigenstates $\psi_{n,\bf{k}}(\bf{r})$ associated with the eigenvalue $E_{n,\bf{k}}$, namely $\tau\mathcal{O} \psi_{n,\bf{k}}(\bf{r})=$ $\exp{(-i\vec{\tau}\cdot\bf{k})}\psi_{n,\mathcal{O}\bf{k}}(\bf{r})$ \cite{dresselhaus2007group}. 
Since the Hamiltonian is invariant under $\tau \mathcal{O}$, $E_{n,{\bf k}}=E_{n,\mathcal{O}{\bf k}}$.

Let us examine, for example, $\Omega^{L_{\eta}}_{yx,n}({\bf k})$. Inserting the identity $(\tau\mathcal{O})^{\dagger}(\tau\mathcal{O})=\mathbb{1}$ into the orbital-weighted Berry curvature and using the above relations we obtain
\begin{widetext}
\begin{eqnarray} 
\Omega^{L_{\eta}}_{yx,n}({\bf k})&=&2\hbar \sum_{m\neq n}\text{Im} \left[ \frac{\langle u_{n,{\bf k}}\big| (\tau\mathcal{O})^{\dagger} (\tau\mathcal{O}) j_{y,{\bf k}}^{L_{\eta}} (\tau\mathcal{O})^{\dagger} (\tau\mathcal{O})\big|u_{m,{\bf k}}\rangle \langle u_{m,{\bf k}}\big|(\tau\mathcal{O})^{\dagger} (\tau\mathcal{O}) v_{x,{\bf k}}(\tau\mathcal{O})^{\dagger} (\tau\mathcal{O})\big|u_{n,{\bf k}}\rangle}{(E_{n,{\bf k}}-E_{m,{\bf k}}+i0^+)^2}\right].
\end{eqnarray}
The restrictions on the conductivity tensor depend on how the Cartesian components of the velocity and angular momentum operators transform under the group symmetry operations. This information is contained in its character table, which shows that, for the point group of phosphorene, they only acquire a sign $s_{\mathcal{O},\hat{A }}=\pm 1$ \cite{Symmetry-Phosphorene-PhysRevB.90.115439} under such operations, as table \ref{tableSymSign} illustrates. Therefore,
\begin{eqnarray} 
\Omega^{L_{\eta}}_{yx,n}({\bf k})&=&2\hbar \sum_{m\neq n}\text{Im} \left[ \frac{\langle u_{n,\mathcal{O}{\bf k}}\big| s_{\mathcal{O},\hat{v}_{y}} s_{\mathcal{O},\hat{L}_{\eta}} j_{y,\mathcal{O}{\bf k}}^{L_{\eta}} \big|u_{m, \mathcal{O}{\bf k}}\rangle \langle u_{m, \mathcal{O}{\bf k}}\big| s_{\mathcal{O},\hat{v}_{x}} v_{x,\mathcal{O}{\bf k}}\big|u_{n,\mathcal{O}{\bf k}}\rangle}{(E_{n,\mathcal{O}{\bf k}}-E_{m,\mathcal{O}{\bf k}}+i0^+)^2}\right] \nonumber \\
&=&s_{\mathcal{O},\hat{v}_{x}} s_{\mathcal{O},\hat{v}_{y}} s_{\mathcal{O},\hat{L}_{\eta}} \Omega^{L_{\eta}}_{yx,n}(\mathcal{O}{\bf k}).
\end{eqnarray}
The same expression holds for $\Omega^{L_{\eta}}_{xy,n}({\bf k})$.
\end{widetext}  
Since $\int d^2{\bf k}=\int d^2(\mathcal{O} {\bf k})$, it follows from Eq. (\ref{conductivity}) that
\begin{eqnarray}
\mathcal{O}: \ \ \sigma^{L_{\eta}}_{{\rm OH}}= \bar{s}^{\eta}_{{\rm OH}}(\mathcal{O}) \sigma^{L_{\eta}}_{{\rm OH}}, \label{ConstrO}
\end{eqnarray}
where $\bar{s}^{\eta}_{{\rm OH}}(\mathcal{O})=s_{\mathcal{O}, v_{x}} \times s_{\mathcal{O}, v_{y}} \times s_{\mathcal{O}, L_{\eta}}$. If $\bar{s}^{\eta}_{{\rm OH}}(\mathcal{O})=+1$ the symmetry $\mathcal{O}$ does not impose a constraint to the OH conductivity. However, if $\bar{s}^{\eta}_{{\rm OH}}(\mathcal{O})=-1$, $\sigma^{L_{\eta}}_{\rm OH}= 0$.

\begin{table}[h!]
	\centering
	\begin{tabular}{||c ||c c | c c | c c | c c | c c | c c ||} 
		\hline
		Symmetry & & $s_{\mathcal{O}, v_{x}}$ & &  $s_{\mathcal{O}, v_{y}}$ & &  $s_{\mathcal{O}, L_{x}}$ & & $s_{\mathcal{O}, L_{y}}$ & &  $s_{\mathcal{O}, L_{z}}$ \\ [0.5ex] 
		\hline
		\hline
		$\mathcal{P}$ & & $-1$ & & $-1$ & & $+1$ & & $+1$ & & $+1$\\ \hline
		$\mathcal{M}_z$ & & $+1$ & & $+1$ & & $-1$ & & $-1$ & & $+1$ \\ \hline
		$\mathcal{M}_x$ & & $-1$ & & $+1$ & & $+1$ & & $-1$ & & $-1$  \\ \hline
		$\mathcal{M}_y$ & & $+1$ & & $-1$ & & $-1$ & & $+1$ & & $-1$ \\ \hline
		$\mathcal{C}_{2z}$ & & $-1$ & & $-1$ & & $-1$ & & $-1$ & & $+1$ \\ \hline
		$\mathcal{C}_{2x}$ & & $+1$ & & $-1$ & & $+1$ & & $-1$ & & $-1$ \\ \hline
		$\mathcal{C}_{2y}$ & & $-1$ & & $+1$ & & $-1$ & & +1 & & $-1$ \\ [0.5ex]
		 \hline
	\end{tabular}
	\vskip 2.5 pt
	\caption{Signs of the velocity and OAM operator component acquired when transformed under the symmorphic part $\mathcal{O}$ of operations associated with the $D_{2h}$ point group of phosphorene. This table can be used to determine the allowed components of the OHC tensor presented in table \ref{tableSym}.}
	\label{tableSymSign} 
\end{table}

%\bibliography{Bibliografia-Phosphorene}

\begin{thebibliography}{82}%
\makeatletter
\providecommand \@ifxundefined [1]{%
 \@ifx{#1\undefined}
}%
\providecommand \@ifnum [1]{%
 \ifnum #1\expandafter \@firstoftwo
 \else \expandafter \@secondoftwo
 \fi
}%
\providecommand \@ifx [1]{%
 \ifx #1\expandafter \@firstoftwo
 \else \expandafter \@secondoftwo
 \fi
}%
\providecommand \natexlab [1]{#1}%
\providecommand \enquote  [1]{``#1''}%
\providecommand \bibnamefont  [1]{#1}%
\providecommand \bibfnamefont [1]{#1}%
\providecommand \citenamefont [1]{#1}%
\providecommand \href@noop [0]{\@secondoftwo}%
\providecommand \href [0]{\begingroup \@sanitize@url \@href}%
\providecommand \@href[1]{\@@startlink{#1}\@@href}%
\providecommand \@@href[1]{\endgroup#1\@@endlink}%
\providecommand \@sanitize@url [0]{\catcode `\\12\catcode `\$12\catcode
  `\&12\catcode `\#12\catcode `\^12\catcode `\_12\catcode `\%12\relax}%
\providecommand \@@startlink[1]{}%
\providecommand \@@endlink[0]{}%
\providecommand \url  [0]{\begingroup\@sanitize@url \@url }%
\providecommand \@url [1]{\endgroup\@href {#1}{\urlprefix }}%
\providecommand \urlprefix  [0]{URL }%
\providecommand \Eprint [0]{\href }%
\providecommand \doibase [0]{https://doi.org/}%
\providecommand \selectlanguage [0]{\@gobble}%
\providecommand \bibinfo  [0]{\@secondoftwo}%
\providecommand \bibfield  [0]{\@secondoftwo}%
\providecommand \translation [1]{[#1]}%
\providecommand \BibitemOpen [0]{}%
\providecommand \bibitemStop [0]{}%
\providecommand \bibitemNoStop [0]{.\EOS\space}%
\providecommand \EOS [0]{\spacefactor3000\relax}%
\providecommand \BibitemShut  [1]{\csname bibitem#1\endcsname}%
\let\auto@bib@innerbib\@empty
%</preamble>
\bibitem [{\citenamefont {Bernevig}\ \emph {et~al.}(2005)\citenamefont
  {Bernevig}, \citenamefont {Hughes},\ and\ \citenamefont
  {Zhang}}]{Bernevig-Hughes-Zhang-PhysRevLett.95.066601}%
  \BibitemOpen
  \bibfield  {author} {\bibinfo {author} {\bibfnamefont {B.~A.}\ \bibnamefont
  {Bernevig}}, \bibinfo {author} {\bibfnamefont {T.~L.}\ \bibnamefont
  {Hughes}},\ and\ \bibinfo {author} {\bibfnamefont {S.-C.}\ \bibnamefont
  {Zhang}},\ }\bibfield  {title} {\bibinfo {title} {Orbitronics: The intrinsic
  orbital current in $p$-doped silicon},\ }\href
  {https://doi.org/10.1103/PhysRevLett.95.066601} {\bibfield  {journal}
  {\bibinfo  {journal} {Phys. Rev. Lett.}\ }\textbf {\bibinfo {volume} {95}},\
  \bibinfo {pages} {066601} (\bibinfo {year} {2005})}\BibitemShut {NoStop}%
\bibitem [{\citenamefont {Phong}\ \emph {et~al.}(2019)\citenamefont {Phong},
  \citenamefont {Addison}, \citenamefont {Ahn}, \citenamefont {Min},
  \citenamefont {Agarwal},\ and\ \citenamefont
  {Mele}}]{Mele-PhysRevLett.123.236403}%
  \BibitemOpen
  \bibfield  {author} {\bibinfo {author} {\bibfnamefont {V.~o.~T.}\
  \bibnamefont {Phong}}, \bibinfo {author} {\bibfnamefont {Z.}~\bibnamefont
  {Addison}}, \bibinfo {author} {\bibfnamefont {S.}~\bibnamefont {Ahn}},
  \bibinfo {author} {\bibfnamefont {H.}~\bibnamefont {Min}}, \bibinfo {author}
  {\bibfnamefont {R.}~\bibnamefont {Agarwal}},\ and\ \bibinfo {author}
  {\bibfnamefont {E.~J.}\ \bibnamefont {Mele}},\ }\bibfield  {title} {\bibinfo
  {title} {Optically controlled orbitronics on a triangular lattice},\ }\href
  {https://doi.org/10.1103/PhysRevLett.123.236403} {\bibfield  {journal}
  {\bibinfo  {journal} {Phys. Rev. Lett.}\ }\textbf {\bibinfo {volume} {123}},\
  \bibinfo {pages} {236403} (\bibinfo {year} {2019})}\BibitemShut {NoStop}%
\bibitem [{\citenamefont {Salemi}\ and\ \citenamefont
  {Oppeneer}(2022{\natexlab{a}})}]{Oppeneer-PhysRevMaterials.6.095001}%
  \BibitemOpen
  \bibfield  {author} {\bibinfo {author} {\bibfnamefont {L.}~\bibnamefont
  {Salemi}}\ and\ \bibinfo {author} {\bibfnamefont {P.~M.}\ \bibnamefont
  {Oppeneer}},\ }\bibfield  {title} {\bibinfo {title} {First-principles theory
  of intrinsic spin and orbital hall and nernst effects in metallic monoatomic
  crystals},\ }\href {https://doi.org/10.1103/PhysRevMaterials.6.095001}
  {\bibfield  {journal} {\bibinfo  {journal} {Phys. Rev. Mater.}\ }\textbf
  {\bibinfo {volume} {6}},\ \bibinfo {pages} {095001} (\bibinfo {year}
  {2022}{\natexlab{a}})}\BibitemShut {NoStop}%
\bibitem [{\citenamefont {Salemi}\ and\ \citenamefont
  {Oppeneer}(2022{\natexlab{b}})}]{Salemi-PhysRevB.106.024410}%
  \BibitemOpen
  \bibfield  {author} {\bibinfo {author} {\bibfnamefont {L.}~\bibnamefont
  {Salemi}}\ and\ \bibinfo {author} {\bibfnamefont {P.~M.}\ \bibnamefont
  {Oppeneer}},\ }\bibfield  {title} {\bibinfo {title} {Theory of magnetic spin
  and orbital hall and nernst effects in bulk ferromagnets},\ }\href
  {https://doi.org/10.1103/PhysRevB.106.024410} {\bibfield  {journal} {\bibinfo
   {journal} {Phys. Rev. B}\ }\textbf {\bibinfo {volume} {106}},\ \bibinfo
  {pages} {024410} (\bibinfo {year} {2022}{\natexlab{b}})}\BibitemShut
  {NoStop}%
\bibitem [{\citenamefont {Bose}\ \emph {et~al.}(2023)\citenamefont {Bose},
  \citenamefont {Kammerbauer}, \citenamefont {Gupta}, \citenamefont {Go},
  \citenamefont {Mokrousov}, \citenamefont {Jakob},\ and\ \citenamefont
  {Kl\"aui}}]{OH-Torque-PhysRevB.107.134423}%
  \BibitemOpen
  \bibfield  {author} {\bibinfo {author} {\bibfnamefont {A.}~\bibnamefont
  {Bose}}, \bibinfo {author} {\bibfnamefont {F.}~\bibnamefont {Kammerbauer}},
  \bibinfo {author} {\bibfnamefont {R.}~\bibnamefont {Gupta}}, \bibinfo
  {author} {\bibfnamefont {D.}~\bibnamefont {Go}}, \bibinfo {author}
  {\bibfnamefont {Y.}~\bibnamefont {Mokrousov}}, \bibinfo {author}
  {\bibfnamefont {G.}~\bibnamefont {Jakob}},\ and\ \bibinfo {author}
  {\bibfnamefont {M.}~\bibnamefont {Kl\"aui}},\ }\bibfield  {title} {\bibinfo
  {title} {Detection of long-range orbital-hall torques},\ }\href
  {https://doi.org/10.1103/PhysRevB.107.134423} {\bibfield  {journal} {\bibinfo
   {journal} {Phys. Rev. B}\ }\textbf {\bibinfo {volume} {107}},\ \bibinfo
  {pages} {134423} (\bibinfo {year} {2023})}\BibitemShut {NoStop}%
\bibitem [{\citenamefont {Go}\ \emph {et~al.}(2023)\citenamefont {Go},
  \citenamefont {An}, \citenamefont {Lee},\ and\ \citenamefont
  {Kim}}]{go2023intrinsic}%
  \BibitemOpen
  \bibfield  {author} {\bibinfo {author} {\bibfnamefont {G.}~\bibnamefont
  {Go}}, \bibinfo {author} {\bibfnamefont {D.}~\bibnamefont {An}}, \bibinfo
  {author} {\bibfnamefont {H.-W.}\ \bibnamefont {Lee}},\ and\ \bibinfo {author}
  {\bibfnamefont {S.~K.}\ \bibnamefont {Kim}},\ }\href@noop {} {\bibinfo
  {title} {Intrinsic magnon orbital hall effect in honeycomb antiferromagnets}}
  (\bibinfo {year} {2023}),\ \Eprint {https://arxiv.org/abs/2303.11687}
  {arXiv:2303.11687 [cond-mat.mes-hall]} \BibitemShut {NoStop}%
\bibitem [{\citenamefont {Zeer}\ \emph {et~al.}(2022)\citenamefont {Zeer},
  \citenamefont {Go}, \citenamefont {Carbone}, \citenamefont {Saunderson},
  \citenamefont {Redies}, \citenamefont {Kl\"aui}, \citenamefont {Ghabboun},
  \citenamefont {Wulfhekel}, \citenamefont {Bl\"ugel},\ and\ \citenamefont
  {Mokrousov}}]{Mokrousov-PhysRevMaterials.6.074004}%
  \BibitemOpen
  \bibfield  {author} {\bibinfo {author} {\bibfnamefont {M.}~\bibnamefont
  {Zeer}}, \bibinfo {author} {\bibfnamefont {D.}~\bibnamefont {Go}}, \bibinfo
  {author} {\bibfnamefont {J.~P.}\ \bibnamefont {Carbone}}, \bibinfo {author}
  {\bibfnamefont {T.~G.}\ \bibnamefont {Saunderson}}, \bibinfo {author}
  {\bibfnamefont {M.}~\bibnamefont {Redies}}, \bibinfo {author} {\bibfnamefont
  {M.}~\bibnamefont {Kl\"aui}}, \bibinfo {author} {\bibfnamefont
  {J.}~\bibnamefont {Ghabboun}}, \bibinfo {author} {\bibfnamefont
  {W.}~\bibnamefont {Wulfhekel}}, \bibinfo {author} {\bibfnamefont
  {S.}~\bibnamefont {Bl\"ugel}},\ and\ \bibinfo {author} {\bibfnamefont
  {Y.}~\bibnamefont {Mokrousov}},\ }\bibfield  {title} {\bibinfo {title} {Spin
  and orbital transport in rare-earth dichalcogenides: The case of
  ${\mathrm{eus}}_{2}$},\ }\href
  {https://doi.org/10.1103/PhysRevMaterials.6.074004} {\bibfield  {journal}
  {\bibinfo  {journal} {Phys. Rev. Mater.}\ }\textbf {\bibinfo {volume} {6}},\
  \bibinfo {pages} {074004} (\bibinfo {year} {2022})}\BibitemShut {NoStop}%
\bibitem [{\citenamefont {Han}\ \emph {et~al.}(2022)\citenamefont {Han},
  \citenamefont {Lee},\ and\ \citenamefont
  {Kim}}]{HW-Lee-PhysRevLett.128.176601}%
  \BibitemOpen
  \bibfield  {author} {\bibinfo {author} {\bibfnamefont {S.}~\bibnamefont
  {Han}}, \bibinfo {author} {\bibfnamefont {H.-W.}\ \bibnamefont {Lee}},\ and\
  \bibinfo {author} {\bibfnamefont {K.-W.}\ \bibnamefont {Kim}},\ }\bibfield
  {title} {\bibinfo {title} {Orbital dynamics in centrosymmetric systems},\
  }\href {https://doi.org/10.1103/PhysRevLett.128.176601} {\bibfield  {journal}
  {\bibinfo  {journal} {Phys. Rev. Lett.}\ }\textbf {\bibinfo {volume} {128}},\
  \bibinfo {pages} {176601} (\bibinfo {year} {2022})}\BibitemShut {NoStop}%
\bibitem [{\citenamefont {Fonseca}\ \emph {et~al.}(2023)\citenamefont
  {Fonseca}, \citenamefont {Pereira},\ and\ \citenamefont
  {Barbosa}}]{fonseca2023orbital}%
  \BibitemOpen
  \bibfield  {author} {\bibinfo {author} {\bibfnamefont {D.~B.}\ \bibnamefont
  {Fonseca}}, \bibinfo {author} {\bibfnamefont {L.~L.~A.}\ \bibnamefont
  {Pereira}},\ and\ \bibinfo {author} {\bibfnamefont {A.~L.~R.}\ \bibnamefont
  {Barbosa}},\ }\href@noop {} {\bibinfo {title} {Orbital hall effect in
  mesoscopic devices}} (\bibinfo {year} {2023}),\ \Eprint
  {https://arxiv.org/abs/2305.01640} {arXiv:2305.01640 [cond-mat.mes-hall]}
  \BibitemShut {NoStop}%
\bibitem [{\citenamefont {Sala}\ and\ \citenamefont
  {Gambardella}(2022)}]{Gambardella-PhysRevResearch.4.033037}%
  \BibitemOpen
  \bibfield  {author} {\bibinfo {author} {\bibfnamefont {G.}~\bibnamefont
  {Sala}}\ and\ \bibinfo {author} {\bibfnamefont {P.}~\bibnamefont
  {Gambardella}},\ }\bibfield  {title} {\bibinfo {title} {Giant orbital hall
  effect and orbital-to-spin conversion in $3d$, $5d$, and $4f$ metallic
  heterostructures},\ }\href {https://doi.org/10.1103/PhysRevResearch.4.033037}
  {\bibfield  {journal} {\bibinfo  {journal} {Phys. Rev. Res.}\ }\textbf
  {\bibinfo {volume} {4}},\ \bibinfo {pages} {033037} (\bibinfo {year}
  {2022})}\BibitemShut {NoStop}%
\bibitem [{\citenamefont {Urazhdin}(2023)}]{urazhdin2023symmetry}%
  \BibitemOpen
  \bibfield  {author} {\bibinfo {author} {\bibfnamefont {S.}~\bibnamefont
  {Urazhdin}},\ }\href@noop {} {\bibinfo {title} {Symmetry constraints on the
  orbital transport in solids}} (\bibinfo {year} {2023}),\ \Eprint
  {https://arxiv.org/abs/2309.04442} {arXiv:2309.04442 [cond-mat.mtrl-sci]}
  \BibitemShut {NoStop}%
\bibitem [{\citenamefont {Busch}\ \emph {et~al.}(2023)\citenamefont {Busch},
  \citenamefont {Mertig},\ and\ \citenamefont {Göbel}}]{busch2023orbital}%
  \BibitemOpen
  \bibfield  {author} {\bibinfo {author} {\bibfnamefont {O.}~\bibnamefont
  {Busch}}, \bibinfo {author} {\bibfnamefont {I.}~\bibnamefont {Mertig}},\ and\
  \bibinfo {author} {\bibfnamefont {B.}~\bibnamefont {Göbel}},\ }\href@noop {}
  {\bibinfo {title} {Orbital hall effect and orbital edge states caused by s
  electrons}} (\bibinfo {year} {2023}),\ \Eprint
  {https://arxiv.org/abs/2306.17295} {arXiv:2306.17295 [cond-mat.mes-hall]}
  \BibitemShut {NoStop}%
\bibitem [{\citenamefont {M.}\ \emph {et~al.}(2023)\citenamefont {M.},
  \citenamefont {Henk}, \citenamefont {Mertig},\ and\ \citenamefont
  {Johansson}}]{m2023spin}%
  \BibitemOpen
  \bibfield  {author} {\bibinfo {author} {\bibfnamefont {S.~L.}\ \bibnamefont
  {M.}}, \bibinfo {author} {\bibfnamefont {J.}~\bibnamefont {Henk}}, \bibinfo
  {author} {\bibfnamefont {I.}~\bibnamefont {Mertig}},\ and\ \bibinfo {author}
  {\bibfnamefont {A.}~\bibnamefont {Johansson}},\ }\href@noop {} {\bibinfo
  {title} {Spin and orbital edelstein effect in a bilayer system with rashba
  interaction}} (\bibinfo {year} {2023}),\ \Eprint
  {https://arxiv.org/abs/2307.02872} {arXiv:2307.02872 [cond-mat.mes-hall]}
  \BibitemShut {NoStop}%
\bibitem [{\citenamefont {Go}\ \emph {et~al.}(2018)\citenamefont {Go},
  \citenamefont {Jo}, \citenamefont {Kim},\ and\ \citenamefont
  {Lee}}]{Go-Hyun-Woo-PhysRevLett.121.086602}%
  \BibitemOpen
  \bibfield  {author} {\bibinfo {author} {\bibfnamefont {D.}~\bibnamefont
  {Go}}, \bibinfo {author} {\bibfnamefont {D.}~\bibnamefont {Jo}}, \bibinfo
  {author} {\bibfnamefont {C.}~\bibnamefont {Kim}},\ and\ \bibinfo {author}
  {\bibfnamefont {H.-W.}\ \bibnamefont {Lee}},\ }\bibfield  {title} {\bibinfo
  {title} {Intrinsic spin and orbital hall effects from orbital texture},\
  }\href {https://doi.org/10.1103/PhysRevLett.121.086602} {\bibfield  {journal}
  {\bibinfo  {journal} {Phys. Rev. Lett.}\ }\textbf {\bibinfo {volume} {121}},\
  \bibinfo {pages} {086602} (\bibinfo {year} {2018})}\BibitemShut {NoStop}%
\bibitem [{\citenamefont {Go}\ \emph {et~al.}(2021)\citenamefont {Go},
  \citenamefont {Jo}, \citenamefont {Lee}, \citenamefont {Kläui},\ and\
  \citenamefont {Mokrousov}}]{Go_EPL-Review}%
  \BibitemOpen
  \bibfield  {author} {\bibinfo {author} {\bibfnamefont {D.}~\bibnamefont
  {Go}}, \bibinfo {author} {\bibfnamefont {D.}~\bibnamefont {Jo}}, \bibinfo
  {author} {\bibfnamefont {H.-W.}\ \bibnamefont {Lee}}, \bibinfo {author}
  {\bibfnamefont {M.}~\bibnamefont {Kläui}},\ and\ \bibinfo {author}
  {\bibfnamefont {Y.}~\bibnamefont {Mokrousov}},\ }\bibfield  {title} {\bibinfo
  {title} {Orbitronics: Orbital currents in solids},\ }\href
  {https://doi.org/10.1209/0295-5075/ac2653} {\bibfield  {journal} {\bibinfo
  {journal} {Europhysics Letters}\ }\textbf {\bibinfo {volume} {135}},\
  \bibinfo {pages} {37001} (\bibinfo {year} {2021})}\BibitemShut {NoStop}%
\bibitem [{\citenamefont {Choi}\ \emph {et~al.}(2023)\citenamefont {Choi},
  \citenamefont {Jo}, \citenamefont {Ko}, \citenamefont {Go}, \citenamefont
  {Kim}, \citenamefont {Park}, \citenamefont {Kim}, \citenamefont {Min},
  \citenamefont {Choi},\ and\ \citenamefont
  {Lee}}]{Go-Experiment-https://doi.org/10.48550/arxiv.2109.14847}%
  \BibitemOpen
  \bibfield  {author} {\bibinfo {author} {\bibfnamefont {Y.-G.}\ \bibnamefont
  {Choi}}, \bibinfo {author} {\bibfnamefont {D.}~\bibnamefont {Jo}}, \bibinfo
  {author} {\bibfnamefont {K.-H.}\ \bibnamefont {Ko}}, \bibinfo {author}
  {\bibfnamefont {D.}~\bibnamefont {Go}}, \bibinfo {author} {\bibfnamefont
  {K.-H.}\ \bibnamefont {Kim}}, \bibinfo {author} {\bibfnamefont {H.~G.}\
  \bibnamefont {Park}}, \bibinfo {author} {\bibfnamefont {C.}~\bibnamefont
  {Kim}}, \bibinfo {author} {\bibfnamefont {B.-C.}\ \bibnamefont {Min}},
  \bibinfo {author} {\bibfnamefont {G.-M.}\ \bibnamefont {Choi}},\ and\
  \bibinfo {author} {\bibfnamefont {H.-W.}\ \bibnamefont {Lee}},\ }\bibfield
  {title} {\bibinfo {title} {Observation of the orbital hall effect in a light
  metal ti},\ }\href {https://doi.org/10.1038/s41586-023-06101-9} {\bibfield
  {journal} {\bibinfo  {journal} {Nature}\ }\textbf {\bibinfo {volume} {619}},\
  \bibinfo {pages} {52} (\bibinfo {year} {2023})}\BibitemShut {NoStop}%
\bibitem [{\citenamefont {Zheng}\ \emph {et~al.}(2020)\citenamefont {Zheng},
  \citenamefont {Guo}, \citenamefont {Jo}, \citenamefont {Go}, \citenamefont
  {Wang}, \citenamefont {Chen}, \citenamefont {Yin}, \citenamefont {Wang},
  \citenamefont {Yu}, \citenamefont {He}, \citenamefont {Lee}, \citenamefont
  {Teng},\ and\ \citenamefont
  {Zhu}}]{Zheng-OrbTorque-PhysRevResearch.2.013127}%
  \BibitemOpen
  \bibfield  {author} {\bibinfo {author} {\bibfnamefont {Z.~C.}\ \bibnamefont
  {Zheng}}, \bibinfo {author} {\bibfnamefont {Q.~X.}\ \bibnamefont {Guo}},
  \bibinfo {author} {\bibfnamefont {D.}~\bibnamefont {Jo}}, \bibinfo {author}
  {\bibfnamefont {D.}~\bibnamefont {Go}}, \bibinfo {author} {\bibfnamefont
  {L.~H.}\ \bibnamefont {Wang}}, \bibinfo {author} {\bibfnamefont {H.~C.}\
  \bibnamefont {Chen}}, \bibinfo {author} {\bibfnamefont {W.}~\bibnamefont
  {Yin}}, \bibinfo {author} {\bibfnamefont {X.~M.}\ \bibnamefont {Wang}},
  \bibinfo {author} {\bibfnamefont {G.~H.}\ \bibnamefont {Yu}}, \bibinfo
  {author} {\bibfnamefont {W.}~\bibnamefont {He}}, \bibinfo {author}
  {\bibfnamefont {H.-W.}\ \bibnamefont {Lee}}, \bibinfo {author} {\bibfnamefont
  {J.}~\bibnamefont {Teng}},\ and\ \bibinfo {author} {\bibfnamefont
  {T.}~\bibnamefont {Zhu}},\ }\bibfield  {title} {\bibinfo {title}
  {Magnetization switching driven by current-induced torque from weakly
  spin-orbit coupled zr},\ }\href
  {https://doi.org/10.1103/PhysRevResearch.2.013127} {\bibfield  {journal}
  {\bibinfo  {journal} {Phys. Rev. Res.}\ }\textbf {\bibinfo {volume} {2}},\
  \bibinfo {pages} {013127} (\bibinfo {year} {2020})}\BibitemShut {NoStop}%
\bibitem [{\citenamefont {Lee}\ \emph {et~al.}(2021{\natexlab{a}})\citenamefont
  {Lee}, \citenamefont {Kang}, \citenamefont {Go}, \citenamefont {Kim},
  \citenamefont {Kang}, \citenamefont {Lee}, \citenamefont {Lee}, \citenamefont
  {Kang}, \citenamefont {Lee}, \citenamefont {Mokrousov}, \citenamefont {Kim},
  \citenamefont {Kim}, \citenamefont {Lee},\ and\ \citenamefont
  {Park}}]{Lee-OrbTorque-10.1038/s42005-021-00737-7}%
  \BibitemOpen
  \bibfield  {author} {\bibinfo {author} {\bibfnamefont {S.}~\bibnamefont
  {Lee}}, \bibinfo {author} {\bibfnamefont {M.-G.}\ \bibnamefont {Kang}},
  \bibinfo {author} {\bibfnamefont {D.}~\bibnamefont {Go}}, \bibinfo {author}
  {\bibfnamefont {D.}~\bibnamefont {Kim}}, \bibinfo {author} {\bibfnamefont
  {J.-H.}\ \bibnamefont {Kang}}, \bibinfo {author} {\bibfnamefont
  {T.}~\bibnamefont {Lee}}, \bibinfo {author} {\bibfnamefont {G.-H.}\
  \bibnamefont {Lee}}, \bibinfo {author} {\bibfnamefont {J.}~\bibnamefont
  {Kang}}, \bibinfo {author} {\bibfnamefont {N.~J.}\ \bibnamefont {Lee}},
  \bibinfo {author} {\bibfnamefont {Y.}~\bibnamefont {Mokrousov}}, \bibinfo
  {author} {\bibfnamefont {S.}~\bibnamefont {Kim}}, \bibinfo {author}
  {\bibfnamefont {K.-J.}\ \bibnamefont {Kim}}, \bibinfo {author} {\bibfnamefont
  {K.-J.}\ \bibnamefont {Lee}},\ and\ \bibinfo {author} {\bibfnamefont {B.-G.}\
  \bibnamefont {Park}},\ }\bibfield  {title} {\bibinfo {title} {Efficient
  conversion of orbital hall current to spin current for spin-orbit torque
  switching},\ }\bibfield  {journal} {\bibinfo  {journal} {Communications
  Physics}\ }\textbf {\bibinfo {volume} {4}},\ \href
  {https://doi.org/10.1038/s42005-021-00737-7} {10.1038/s42005-021-00737-7}
  (\bibinfo {year} {2021}{\natexlab{a}})\BibitemShut {NoStop}%
\bibitem [{\citenamefont {Lee}\ \emph {et~al.}(2021{\natexlab{b}})\citenamefont
  {Lee}, \citenamefont {Go}, \citenamefont {Park}, \citenamefont {Jeong},
  \citenamefont {Ko}, \citenamefont {Yun}, \citenamefont {Jo}, \citenamefont
  {Lee}, \citenamefont {Go}, \citenamefont {Oh}, \citenamefont {Kim},
  \citenamefont {Park}, \citenamefont {Min}, \citenamefont {Koo}, \citenamefont
  {Lee}, \citenamefont {Lee},\ and\ \citenamefont {Lee}}]{Lee2021}%
  \BibitemOpen
  \bibfield  {author} {\bibinfo {author} {\bibfnamefont {D.}~\bibnamefont
  {Lee}}, \bibinfo {author} {\bibfnamefont {D.}~\bibnamefont {Go}}, \bibinfo
  {author} {\bibfnamefont {H.-J.}\ \bibnamefont {Park}}, \bibinfo {author}
  {\bibfnamefont {W.}~\bibnamefont {Jeong}}, \bibinfo {author} {\bibfnamefont
  {H.-W.}\ \bibnamefont {Ko}}, \bibinfo {author} {\bibfnamefont
  {D.}~\bibnamefont {Yun}}, \bibinfo {author} {\bibfnamefont {D.}~\bibnamefont
  {Jo}}, \bibinfo {author} {\bibfnamefont {S.}~\bibnamefont {Lee}}, \bibinfo
  {author} {\bibfnamefont {G.}~\bibnamefont {Go}}, \bibinfo {author}
  {\bibfnamefont {J.~H.}\ \bibnamefont {Oh}}, \bibinfo {author} {\bibfnamefont
  {K.-J.}\ \bibnamefont {Kim}}, \bibinfo {author} {\bibfnamefont {B.-G.}\
  \bibnamefont {Park}}, \bibinfo {author} {\bibfnamefont {B.-C.}\ \bibnamefont
  {Min}}, \bibinfo {author} {\bibfnamefont {H.~C.}\ \bibnamefont {Koo}},
  \bibinfo {author} {\bibfnamefont {H.-W.}\ \bibnamefont {Lee}}, \bibinfo
  {author} {\bibfnamefont {O.}~\bibnamefont {Lee}},\ and\ \bibinfo {author}
  {\bibfnamefont {K.-J.}\ \bibnamefont {Lee}},\ }\bibfield  {title} {\bibinfo
  {title} {Orbital torque in magnetic bilayers},\ }\bibfield  {journal}
  {\bibinfo  {journal} {Nature Communications}\ }\textbf {\bibinfo {volume}
  {12}},\ \href {https://doi.org/10.1038/s41467-021-26650-9}
  {10.1038/s41467-021-26650-9} (\bibinfo {year}
  {2021}{\natexlab{b}})\BibitemShut {NoStop}%
\bibitem [{\citenamefont {Canonico}\ \emph
  {et~al.}(2020{\natexlab{a}})\citenamefont {Canonico}, \citenamefont {Cysne},
  \citenamefont {Molina-Sanchez}, \citenamefont {Muniz},\ and\ \citenamefont
  {Rappoport}}]{Canonico-PhysRevB.101.161409}%
  \BibitemOpen
  \bibfield  {author} {\bibinfo {author} {\bibfnamefont {L.~M.}\ \bibnamefont
  {Canonico}}, \bibinfo {author} {\bibfnamefont {T.~P.}\ \bibnamefont {Cysne}},
  \bibinfo {author} {\bibfnamefont {A.}~\bibnamefont {Molina-Sanchez}},
  \bibinfo {author} {\bibfnamefont {R.~B.}\ \bibnamefont {Muniz}},\ and\
  \bibinfo {author} {\bibfnamefont {T.~G.}\ \bibnamefont {Rappoport}},\
  }\bibfield  {title} {\bibinfo {title} {Orbital hall insulating phase in
  transition metal dichalcogenide monolayers},\ }\href
  {https://doi.org/10.1103/PhysRevB.101.161409} {\bibfield  {journal} {\bibinfo
   {journal} {Phys. Rev. B}\ }\textbf {\bibinfo {volume} {101}},\ \bibinfo
  {pages} {161409} (\bibinfo {year} {2020}{\natexlab{a}})}\BibitemShut
  {NoStop}%
\bibitem [{\citenamefont {Canonico}\ \emph
  {et~al.}(2020{\natexlab{b}})\citenamefont {Canonico}, \citenamefont {Cysne},
  \citenamefont {Rappoport},\ and\ \citenamefont
  {Muniz}}]{Canonico-PhysRevB.101.075429}%
  \BibitemOpen
  \bibfield  {author} {\bibinfo {author} {\bibfnamefont {L.~M.}\ \bibnamefont
  {Canonico}}, \bibinfo {author} {\bibfnamefont {T.~P.}\ \bibnamefont {Cysne}},
  \bibinfo {author} {\bibfnamefont {T.~G.}\ \bibnamefont {Rappoport}},\ and\
  \bibinfo {author} {\bibfnamefont {R.~B.}\ \bibnamefont {Muniz}},\ }\bibfield
  {title} {\bibinfo {title} {Two-dimensional orbital hall insulators},\ }\href
  {https://doi.org/10.1103/PhysRevB.101.075429} {\bibfield  {journal} {\bibinfo
   {journal} {Phys. Rev. B}\ }\textbf {\bibinfo {volume} {101}},\ \bibinfo
  {pages} {075429} (\bibinfo {year} {2020}{\natexlab{b}})}\BibitemShut
  {NoStop}%
\bibitem [{\citenamefont {Costa}\ \emph {et~al.}(2023)\citenamefont {Costa},
  \citenamefont {Focassio}, \citenamefont {Canonico}, \citenamefont {Cysne},
  \citenamefont {Schleder}, \citenamefont {Muniz}, \citenamefont {Fazzio},\
  and\ \citenamefont {Rappoport}}]{Costa2023}%
  \BibitemOpen
  \bibfield  {author} {\bibinfo {author} {\bibfnamefont {M.}~\bibnamefont
  {Costa}}, \bibinfo {author} {\bibfnamefont {B.}~\bibnamefont {Focassio}},
  \bibinfo {author} {\bibfnamefont {L.~M.}\ \bibnamefont {Canonico}}, \bibinfo
  {author} {\bibfnamefont {T.~P.}\ \bibnamefont {Cysne}}, \bibinfo {author}
  {\bibfnamefont {G.~R.}\ \bibnamefont {Schleder}}, \bibinfo {author}
  {\bibfnamefont {R.~B.}\ \bibnamefont {Muniz}}, \bibinfo {author}
  {\bibfnamefont {A.}~\bibnamefont {Fazzio}},\ and\ \bibinfo {author}
  {\bibfnamefont {T.~G.}\ \bibnamefont {Rappoport}},\ }\bibfield  {title}
  {\bibinfo {title} {Connecting higher-order topology with the orbital hall
  effect in monolayers of transition metal dichalcogenides},\ }\href
  {https://doi.org/10.1103/PhysRevLett.130.116204} {\bibfield  {journal}
  {\bibinfo  {journal} {Phys. Rev. Lett.}\ }\textbf {\bibinfo {volume} {130}},\
  \bibinfo {pages} {116204} (\bibinfo {year} {2023})}\BibitemShut {NoStop}%
\bibitem [{\citenamefont {Cysne}\ \emph
  {et~al.}(2021{\natexlab{a}})\citenamefont {Cysne}, \citenamefont {Costa},
  \citenamefont {Canonico}, \citenamefont {Nardelli}, \citenamefont {Muniz},\
  and\ \citenamefont {Rappoport}}]{Cysne-PhysRevLett.126.056601}%
  \BibitemOpen
  \bibfield  {author} {\bibinfo {author} {\bibfnamefont {T.~P.}\ \bibnamefont
  {Cysne}}, \bibinfo {author} {\bibfnamefont {M.}~\bibnamefont {Costa}},
  \bibinfo {author} {\bibfnamefont {L.~M.}\ \bibnamefont {Canonico}}, \bibinfo
  {author} {\bibfnamefont {M.~B.}\ \bibnamefont {Nardelli}}, \bibinfo {author}
  {\bibfnamefont {R.~B.}\ \bibnamefont {Muniz}},\ and\ \bibinfo {author}
  {\bibfnamefont {T.~G.}\ \bibnamefont {Rappoport}},\ }\bibfield  {title}
  {\bibinfo {title} {Disentangling orbital and valley hall effects in bilayers
  of transition metal dichalcogenides},\ }\href
  {https://doi.org/10.1103/PhysRevLett.126.056601} {\bibfield  {journal}
  {\bibinfo  {journal} {Phys. Rev. Lett.}\ }\textbf {\bibinfo {volume} {126}},\
  \bibinfo {pages} {056601} (\bibinfo {year} {2021}{\natexlab{a}})}\BibitemShut
  {NoStop}%
\bibitem [{\citenamefont {Cysne}\ \emph {et~al.}(2022)\citenamefont {Cysne},
  \citenamefont {Bhowal}, \citenamefont {Vignale},\ and\ \citenamefont
  {Rappoport}}]{Cysne-PhysRevB.105.195421}%
  \BibitemOpen
  \bibfield  {author} {\bibinfo {author} {\bibfnamefont {T.~P.}\ \bibnamefont
  {Cysne}}, \bibinfo {author} {\bibfnamefont {S.}~\bibnamefont {Bhowal}},
  \bibinfo {author} {\bibfnamefont {G.}~\bibnamefont {Vignale}},\ and\ \bibinfo
  {author} {\bibfnamefont {T.~G.}\ \bibnamefont {Rappoport}},\ }\bibfield
  {title} {\bibinfo {title} {Orbital hall effect in bilayer transition metal
  dichalcogenides: From the intra-atomic approximation to the bloch states
  orbital magnetic moment approach},\ }\href
  {https://doi.org/10.1103/PhysRevB.105.195421} {\bibfield  {journal} {\bibinfo
   {journal} {Phys. Rev. B}\ }\textbf {\bibinfo {volume} {105}},\ \bibinfo
  {pages} {195421} (\bibinfo {year} {2022})}\BibitemShut {NoStop}%
\bibitem [{\citenamefont {Bhowal}\ and\ \citenamefont
  {Vignale}(2021)}]{Bhowal-PhysRevB.103.195309}%
  \BibitemOpen
  \bibfield  {author} {\bibinfo {author} {\bibfnamefont {S.}~\bibnamefont
  {Bhowal}}\ and\ \bibinfo {author} {\bibfnamefont {G.}~\bibnamefont
  {Vignale}},\ }\bibfield  {title} {\bibinfo {title} {Orbital hall effect as an
  alternative to valley hall effect in gapped graphene},\ }\href
  {https://doi.org/10.1103/PhysRevB.103.195309} {\bibfield  {journal} {\bibinfo
   {journal} {Phys. Rev. B}\ }\textbf {\bibinfo {volume} {103}},\ \bibinfo
  {pages} {195309} (\bibinfo {year} {2021})}\BibitemShut {NoStop}%
\bibitem [{\citenamefont {Salvador-Sánchez}\ \emph {et~al.}(2022)\citenamefont
  {Salvador-Sánchez}, \citenamefont {Canonico}, \citenamefont
  {Pérez-Rodríguez}, \citenamefont {Cysne}, \citenamefont {Baba},
  \citenamefont {Clericò}, \citenamefont {Vila}, \citenamefont {Vaquero},
  \citenamefont {Delgado-Notario}, \citenamefont {Caridad}, \citenamefont
  {Watanabe}, \citenamefont {Taniguchi}, \citenamefont {Molina}, \citenamefont
  {Domínguez-Adame}, \citenamefont {Roche}, \citenamefont {Diez},
  \citenamefont {Rappoport},\ and\ \citenamefont
  {Amado}}]{Salvador-Sanchez-https://doi.org/10.48550/arxiv.2206.04565}%
  \BibitemOpen
  \bibfield  {author} {\bibinfo {author} {\bibfnamefont {J.}~\bibnamefont
  {Salvador-Sánchez}}, \bibinfo {author} {\bibfnamefont {L.~M.}\ \bibnamefont
  {Canonico}}, \bibinfo {author} {\bibfnamefont {A.}~\bibnamefont
  {Pérez-Rodríguez}}, \bibinfo {author} {\bibfnamefont {T.~P.}\ \bibnamefont
  {Cysne}}, \bibinfo {author} {\bibfnamefont {Y.}~\bibnamefont {Baba}},
  \bibinfo {author} {\bibfnamefont {V.}~\bibnamefont {Clericò}}, \bibinfo
  {author} {\bibfnamefont {M.}~\bibnamefont {Vila}}, \bibinfo {author}
  {\bibfnamefont {D.}~\bibnamefont {Vaquero}}, \bibinfo {author} {\bibfnamefont
  {J.~A.}\ \bibnamefont {Delgado-Notario}}, \bibinfo {author} {\bibfnamefont
  {J.~M.}\ \bibnamefont {Caridad}}, \bibinfo {author} {\bibfnamefont
  {K.}~\bibnamefont {Watanabe}}, \bibinfo {author} {\bibfnamefont
  {T.}~\bibnamefont {Taniguchi}}, \bibinfo {author} {\bibfnamefont {R.~A.}\
  \bibnamefont {Molina}}, \bibinfo {author} {\bibfnamefont {F.}~\bibnamefont
  {Domínguez-Adame}}, \bibinfo {author} {\bibfnamefont {S.}~\bibnamefont
  {Roche}}, \bibinfo {author} {\bibfnamefont {E.}~\bibnamefont {Diez}},
  \bibinfo {author} {\bibfnamefont {T.~G.}\ \bibnamefont {Rappoport}},\ and\
  \bibinfo {author} {\bibfnamefont {M.}~\bibnamefont {Amado}},\ }\href
  {https://doi.org/10.48550/ARXIV.2206.04565} {\bibinfo {title} {Generation and
  control of non-local chiral currents in graphene superlattices by orbital
  hall effect}} (\bibinfo {year} {2022})\BibitemShut {NoStop}%
\bibitem [{\citenamefont {Li}\ and\ \citenamefont
  {Appelbaum}(2014)}]{Symmetry-Phosphorene-PhysRevB.90.115439}%
  \BibitemOpen
  \bibfield  {author} {\bibinfo {author} {\bibfnamefont {P.}~\bibnamefont
  {Li}}\ and\ \bibinfo {author} {\bibfnamefont {I.}~\bibnamefont {Appelbaum}},\
  }\bibfield  {title} {\bibinfo {title} {Electrons and holes in phosphorene},\
  }\href {https://doi.org/10.1103/PhysRevB.90.115439} {\bibfield  {journal}
  {\bibinfo  {journal} {Phys. Rev. B}\ }\textbf {\bibinfo {volume} {90}},\
  \bibinfo {pages} {115439} (\bibinfo {year} {2014})}\BibitemShut {NoStop}%
\bibitem [{\citenamefont {Rodin}\ \emph {et~al.}(2014)\citenamefont {Rodin},
  \citenamefont {Carvalho},\ and\ \citenamefont
  {Castro~Neto}}]{Phosphorne-SpectraStrain-PhysRevLett.112.176801}%
  \BibitemOpen
  \bibfield  {author} {\bibinfo {author} {\bibfnamefont {A.~S.}\ \bibnamefont
  {Rodin}}, \bibinfo {author} {\bibfnamefont {A.}~\bibnamefont {Carvalho}},\
  and\ \bibinfo {author} {\bibfnamefont {A.~H.}\ \bibnamefont {Castro~Neto}},\
  }\bibfield  {title} {\bibinfo {title} {Strain-induced gap modification in
  black phosphorus},\ }\href {https://doi.org/10.1103/PhysRevLett.112.176801}
  {\bibfield  {journal} {\bibinfo  {journal} {Phys. Rev. Lett.}\ }\textbf
  {\bibinfo {volume} {112}},\ \bibinfo {pages} {176801} (\bibinfo {year}
  {2014})}\BibitemShut {NoStop}%
\bibitem [{\citenamefont {Taghizadeh~Sisakht}\ \emph
  {et~al.}(2016)\citenamefont {Taghizadeh~Sisakht}, \citenamefont {Fazileh},
  \citenamefont {Zare}, \citenamefont {Zarenia},\ and\ \citenamefont
  {Peeters}}]{Phosphorene-Spectra-strain-PhysRevB.94.085417}%
  \BibitemOpen
  \bibfield  {author} {\bibinfo {author} {\bibfnamefont {E.}~\bibnamefont
  {Taghizadeh~Sisakht}}, \bibinfo {author} {\bibfnamefont {F.}~\bibnamefont
  {Fazileh}}, \bibinfo {author} {\bibfnamefont {M.~H.}\ \bibnamefont {Zare}},
  \bibinfo {author} {\bibfnamefont {M.}~\bibnamefont {Zarenia}},\ and\ \bibinfo
  {author} {\bibfnamefont {F.~M.}\ \bibnamefont {Peeters}},\ }\bibfield
  {title} {\bibinfo {title} {Strain-induced topological phase transition in
  phosphorene and in phosphorene nanoribbons},\ }\href
  {https://doi.org/10.1103/PhysRevB.94.085417} {\bibfield  {journal} {\bibinfo
  {journal} {Phys. Rev. B}\ }\textbf {\bibinfo {volume} {94}},\ \bibinfo
  {pages} {085417} (\bibinfo {year} {2016})}\BibitemShut {NoStop}%
\bibitem [{\citenamefont {Rudenko}\ and\ \citenamefont
  {Katsnelson}(2014)}]{Phosphorene-ModelRudenko-PhysRevB.89.201408}%
  \BibitemOpen
  \bibfield  {author} {\bibinfo {author} {\bibfnamefont {A.~N.}\ \bibnamefont
  {Rudenko}}\ and\ \bibinfo {author} {\bibfnamefont {M.~I.}\ \bibnamefont
  {Katsnelson}},\ }\bibfield  {title} {\bibinfo {title} {Quasiparticle band
  structure and tight-binding model for single- and bilayer black phosphorus},\
  }\href {https://doi.org/10.1103/PhysRevB.89.201408} {\bibfield  {journal}
  {\bibinfo  {journal} {Phys. Rev. B}\ }\textbf {\bibinfo {volume} {89}},\
  \bibinfo {pages} {201408} (\bibinfo {year} {2014})}\BibitemShut {NoStop}%
\bibitem [{\citenamefont {Faria~Junior}\ \emph {et~al.}(2019)\citenamefont
  {Faria~Junior}, \citenamefont {Kurpas}, \citenamefont {Gmitra},\ and\
  \citenamefont {Fabian}}]{Paulo-Fabian-PhysRevB.100.115203}%
  \BibitemOpen
  \bibfield  {author} {\bibinfo {author} {\bibfnamefont {P.~E.}\ \bibnamefont
  {Faria~Junior}}, \bibinfo {author} {\bibfnamefont {M.}~\bibnamefont
  {Kurpas}}, \bibinfo {author} {\bibfnamefont {M.}~\bibnamefont {Gmitra}},\
  and\ \bibinfo {author} {\bibfnamefont {J.}~\bibnamefont {Fabian}},\
  }\bibfield  {title} {\bibinfo {title}
  {$k\ifmmode\cdot\else\textperiodcentered\fi{}p$ theory for phosphorene:
  Effective $g$-factors, landau levels, and excitons},\ }\href
  {https://doi.org/10.1103/PhysRevB.100.115203} {\bibfield  {journal} {\bibinfo
   {journal} {Phys. Rev. B}\ }\textbf {\bibinfo {volume} {100}},\ \bibinfo
  {pages} {115203} (\bibinfo {year} {2019})}\BibitemShut {NoStop}%
\bibitem [{\citenamefont {Liu}\ \emph {et~al.}(2015)\citenamefont {Liu},
  \citenamefont {Zhang}, \citenamefont {Abdalla}, \citenamefont {Fazzio},\ and\
  \citenamefont {Zunger}}]{ElectricField-Fazzio-Zunger-doi:10.1021/nl5043769}%
  \BibitemOpen
  \bibfield  {author} {\bibinfo {author} {\bibfnamefont {Q.}~\bibnamefont
  {Liu}}, \bibinfo {author} {\bibfnamefont {X.}~\bibnamefont {Zhang}}, \bibinfo
  {author} {\bibfnamefont {L.~B.}\ \bibnamefont {Abdalla}}, \bibinfo {author}
  {\bibfnamefont {A.}~\bibnamefont {Fazzio}},\ and\ \bibinfo {author}
  {\bibfnamefont {A.}~\bibnamefont {Zunger}},\ }\bibfield  {title} {\bibinfo
  {title} {Switching a normal insulator into a topological insulator via
  electric field with application to phosphorene},\ }\href
  {https://doi.org/10.1021/nl5043769} {\bibfield  {journal} {\bibinfo
  {journal} {Nano Letters}\ }\textbf {\bibinfo {volume} {15}},\ \bibinfo
  {pages} {1222} (\bibinfo {year} {2015})},\ \bibinfo {note} {pMID: 25607525},\
  \Eprint {https://arxiv.org/abs/https://doi.org/10.1021/nl5043769}
  {https://doi.org/10.1021/nl5043769} \BibitemShut {NoStop}%
\bibitem [{\citenamefont {Popovi\ifmmode~\acute{c}\else \'{c}\fi{}}\ \emph
  {et~al.}(2015)\citenamefont {Popovi\ifmmode~\acute{c}\else \'{c}\fi{}},
  \citenamefont {Kurdestany},\ and\ \citenamefont {Satpathy}}]{Popovic2015}%
  \BibitemOpen
  \bibfield  {author} {\bibinfo {author} {\bibfnamefont {Z.~S.}\ \bibnamefont
  {Popovi\ifmmode~\acute{c}\else \'{c}\fi{}}}, \bibinfo {author} {\bibfnamefont
  {J.~M.}\ \bibnamefont {Kurdestany}},\ and\ \bibinfo {author} {\bibfnamefont
  {S.}~\bibnamefont {Satpathy}},\ }\bibfield  {title} {\bibinfo {title}
  {Electronic structure and anisotropic rashba spin-orbit coupling in monolayer
  black phosphorus},\ }\href {https://doi.org/10.1103/PhysRevB.92.035135}
  {\bibfield  {journal} {\bibinfo  {journal} {Phys. Rev. B}\ }\textbf {\bibinfo
  {volume} {92}},\ \bibinfo {pages} {035135} (\bibinfo {year}
  {2015})}\BibitemShut {NoStop}%
\bibitem [{\citenamefont {Avsar}\ \emph {et~al.}(2017)\citenamefont {Avsar},
  \citenamefont {Tan}, \citenamefont {Kurpas}, \citenamefont {Gmitra},
  \citenamefont {Watanabe}, \citenamefont {Taniguchi}, \citenamefont {Fabian},\
  and\ \citenamefont {\"{O}zyilmaz}}]{Avsar2017}%
  \BibitemOpen
  \bibfield  {author} {\bibinfo {author} {\bibfnamefont {A.}~\bibnamefont
  {Avsar}}, \bibinfo {author} {\bibfnamefont {J.~Y.}\ \bibnamefont {Tan}},
  \bibinfo {author} {\bibfnamefont {M.}~\bibnamefont {Kurpas}}, \bibinfo
  {author} {\bibfnamefont {M.}~\bibnamefont {Gmitra}}, \bibinfo {author}
  {\bibfnamefont {K.}~\bibnamefont {Watanabe}}, \bibinfo {author}
  {\bibfnamefont {T.}~\bibnamefont {Taniguchi}}, \bibinfo {author}
  {\bibfnamefont {J.}~\bibnamefont {Fabian}},\ and\ \bibinfo {author}
  {\bibfnamefont {B.}~\bibnamefont {\"{O}zyilmaz}},\ }\bibfield  {title}
  {\bibinfo {title} {Gate-tunable black phosphorus spin valve with nanosecond
  spin lifetimes},\ }\href {https://doi.org/10.1038/nphys4141} {\bibfield
  {journal} {\bibinfo  {journal} {Nature Physics}\ }\textbf {\bibinfo {volume}
  {13}},\ \bibinfo {pages} {888} (\bibinfo {year} {2017})}\BibitemShut
  {NoStop}%
\bibitem [{\citenamefont {Hu}\ \emph {et~al.}(2016)\citenamefont {Hu},
  \citenamefont {Wu}, \citenamefont {Zeng}, \citenamefont {Deng},\ and\
  \citenamefont
  {Kan}}]{Hu-SymmetriesPolarization-doi:10.1021/acs.nanolett.6b04630}%
  \BibitemOpen
  \bibfield  {author} {\bibinfo {author} {\bibfnamefont {T.}~\bibnamefont
  {Hu}}, \bibinfo {author} {\bibfnamefont {H.}~\bibnamefont {Wu}}, \bibinfo
  {author} {\bibfnamefont {H.}~\bibnamefont {Zeng}}, \bibinfo {author}
  {\bibfnamefont {K.}~\bibnamefont {Deng}},\ and\ \bibinfo {author}
  {\bibfnamefont {E.}~\bibnamefont {Kan}},\ }\bibfield  {title} {\bibinfo
  {title} {New ferroelectric phase in atomic-thick phosphorene nanoribbons:
  Existence of in-plane electric polarization},\ }\href
  {https://doi.org/10.1021/acs.nanolett.6b04630} {\bibfield  {journal}
  {\bibinfo  {journal} {Nano Letters}\ }\textbf {\bibinfo {volume} {16}},\
  \bibinfo {pages} {8015} (\bibinfo {year} {2016})},\ \bibinfo {note} {pMID:
  27960526},\ \Eprint
  {https://arxiv.org/abs/https://doi.org/10.1021/acs.nanolett.6b04630}
  {https://doi.org/10.1021/acs.nanolett.6b04630} \BibitemShut {NoStop}%
\bibitem [{\citenamefont {Hohenberg}\ and\ \citenamefont {Kohn}(1964)}]{DFT1}%
  \BibitemOpen
  \bibfield  {author} {\bibinfo {author} {\bibfnamefont {P.}~\bibnamefont
  {Hohenberg}}\ and\ \bibinfo {author} {\bibfnamefont {W.}~\bibnamefont
  {Kohn}},\ }\bibfield  {title} {\bibinfo {title} {Inhomogeneous electron
  gas},\ }\href {https://doi.org/10.1103/PhysRev.136.B864} {\bibfield
  {journal} {\bibinfo  {journal} {Phys. Rev.}\ }\textbf {\bibinfo {volume}
  {136}},\ \bibinfo {pages} {B864} (\bibinfo {year} {1964})}\BibitemShut
  {NoStop}%
\bibitem [{\citenamefont {Kohn}\ and\ \citenamefont {Sham}(1965)}]{DFT2}%
  \BibitemOpen
  \bibfield  {author} {\bibinfo {author} {\bibfnamefont {W.}~\bibnamefont
  {Kohn}}\ and\ \bibinfo {author} {\bibfnamefont {L.~J.}\ \bibnamefont
  {Sham}},\ }\bibfield  {title} {\bibinfo {title} {Self-consistent equations
  including exchange and correlation effects},\ }\href
  {https://doi.org/10.1103/PhysRev.140.A1133} {\bibfield  {journal} {\bibinfo
  {journal} {Phys. Rev.}\ }\textbf {\bibinfo {volume} {140}},\ \bibinfo {pages}
  {A1133} (\bibinfo {year} {1965})}\BibitemShut {NoStop}%
\bibitem [{\citenamefont {Giannozzi}\ \emph {et~al.}(2017)\citenamefont
  {Giannozzi}, \citenamefont {Andreussi}, \citenamefont {Brumme}, \citenamefont
  {Bunau}, \citenamefont {Buongiorno~Nardelli}, \citenamefont {Calandra},
  \citenamefont {Car}, \citenamefont {Cavazzoni}, \citenamefont {Ceresoli},
  \citenamefont {Cococcioni}, \citenamefont {Colonna}, \citenamefont
  {Carnimeo}, \citenamefont {Corso}, \citenamefont {de~Gironcoli},
  \citenamefont {Delugas}, \citenamefont {Jr}, \citenamefont {Ferretti},
  \citenamefont {Floris}, \citenamefont {Fratesi}, \citenamefont {Fugallo},
  \citenamefont {Gebauer}, \citenamefont {Gerstmann}, \citenamefont {Giustino},
  \citenamefont {Gorni}, \citenamefont {Jia}, \citenamefont {Kawamura},
  \citenamefont {Ko}, \citenamefont {Kokalj}, \citenamefont {Küçükbenli},
  \citenamefont {Lazzeri}, \citenamefont {Marsili}, \citenamefont {Marzari},
  \citenamefont {Mauri}, \citenamefont {Nguyen}, \citenamefont {Nguyen},
  \citenamefont {de-la Roza}, \citenamefont {Paulatto}, \citenamefont {Poncé},
  \citenamefont {Rocca}, \citenamefont {Sabatini}, \citenamefont {Santra},
  \citenamefont {Schlipf}, \citenamefont {Seitsonen}, \citenamefont {Smogunov},
  \citenamefont {Timrov}, \citenamefont {Thonhauser}, \citenamefont {Umari},
  \citenamefont {Vast}, \citenamefont {Wu},\ and\ \citenamefont
  {Baroni}}]{QE-2017}%
  \BibitemOpen
  \bibfield  {author} {\bibinfo {author} {\bibfnamefont {P.}~\bibnamefont
  {Giannozzi}}, \bibinfo {author} {\bibfnamefont {O.}~\bibnamefont
  {Andreussi}}, \bibinfo {author} {\bibfnamefont {T.}~\bibnamefont {Brumme}},
  \bibinfo {author} {\bibfnamefont {O.}~\bibnamefont {Bunau}}, \bibinfo
  {author} {\bibfnamefont {M.}~\bibnamefont {Buongiorno~Nardelli}}, \bibinfo
  {author} {\bibfnamefont {M.}~\bibnamefont {Calandra}}, \bibinfo {author}
  {\bibfnamefont {R.}~\bibnamefont {Car}}, \bibinfo {author} {\bibfnamefont
  {C.}~\bibnamefont {Cavazzoni}}, \bibinfo {author} {\bibfnamefont
  {D.}~\bibnamefont {Ceresoli}}, \bibinfo {author} {\bibfnamefont
  {M.}~\bibnamefont {Cococcioni}}, \bibinfo {author} {\bibfnamefont
  {N.}~\bibnamefont {Colonna}}, \bibinfo {author} {\bibfnamefont
  {I.}~\bibnamefont {Carnimeo}}, \bibinfo {author} {\bibfnamefont {A.~D.}\
  \bibnamefont {Corso}}, \bibinfo {author} {\bibfnamefont {S.}~\bibnamefont
  {de~Gironcoli}}, \bibinfo {author} {\bibfnamefont {P.}~\bibnamefont
  {Delugas}}, \bibinfo {author} {\bibfnamefont {R.~A.~D.}\ \bibnamefont {Jr}},
  \bibinfo {author} {\bibfnamefont {A.}~\bibnamefont {Ferretti}}, \bibinfo
  {author} {\bibfnamefont {A.}~\bibnamefont {Floris}}, \bibinfo {author}
  {\bibfnamefont {G.}~\bibnamefont {Fratesi}}, \bibinfo {author} {\bibfnamefont
  {G.}~\bibnamefont {Fugallo}}, \bibinfo {author} {\bibfnamefont
  {R.}~\bibnamefont {Gebauer}}, \bibinfo {author} {\bibfnamefont
  {U.}~\bibnamefont {Gerstmann}}, \bibinfo {author} {\bibfnamefont
  {F.}~\bibnamefont {Giustino}}, \bibinfo {author} {\bibfnamefont
  {T.}~\bibnamefont {Gorni}}, \bibinfo {author} {\bibfnamefont
  {J.}~\bibnamefont {Jia}}, \bibinfo {author} {\bibfnamefont {M.}~\bibnamefont
  {Kawamura}}, \bibinfo {author} {\bibfnamefont {H.-Y.}\ \bibnamefont {Ko}},
  \bibinfo {author} {\bibfnamefont {A.}~\bibnamefont {Kokalj}}, \bibinfo
  {author} {\bibfnamefont {E.}~\bibnamefont {Küçükbenli}}, \bibinfo {author}
  {\bibfnamefont {M.}~\bibnamefont {Lazzeri}}, \bibinfo {author} {\bibfnamefont
  {M.}~\bibnamefont {Marsili}}, \bibinfo {author} {\bibfnamefont
  {N.}~\bibnamefont {Marzari}}, \bibinfo {author} {\bibfnamefont
  {F.}~\bibnamefont {Mauri}}, \bibinfo {author} {\bibfnamefont {N.~L.}\
  \bibnamefont {Nguyen}}, \bibinfo {author} {\bibfnamefont {H.-V.}\
  \bibnamefont {Nguyen}}, \bibinfo {author} {\bibfnamefont {A.~O.}\
  \bibnamefont {de-la Roza}}, \bibinfo {author} {\bibfnamefont
  {L.}~\bibnamefont {Paulatto}}, \bibinfo {author} {\bibfnamefont
  {S.}~\bibnamefont {Poncé}}, \bibinfo {author} {\bibfnamefont
  {D.}~\bibnamefont {Rocca}}, \bibinfo {author} {\bibfnamefont
  {R.}~\bibnamefont {Sabatini}}, \bibinfo {author} {\bibfnamefont
  {B.}~\bibnamefont {Santra}}, \bibinfo {author} {\bibfnamefont
  {M.}~\bibnamefont {Schlipf}}, \bibinfo {author} {\bibfnamefont {A.~P.}\
  \bibnamefont {Seitsonen}}, \bibinfo {author} {\bibfnamefont {A.}~\bibnamefont
  {Smogunov}}, \bibinfo {author} {\bibfnamefont {I.}~\bibnamefont {Timrov}},
  \bibinfo {author} {\bibfnamefont {T.}~\bibnamefont {Thonhauser}}, \bibinfo
  {author} {\bibfnamefont {P.}~\bibnamefont {Umari}}, \bibinfo {author}
  {\bibfnamefont {N.}~\bibnamefont {Vast}}, \bibinfo {author} {\bibfnamefont
  {X.}~\bibnamefont {Wu}},\ and\ \bibinfo {author} {\bibfnamefont
  {S.}~\bibnamefont {Baroni}},\ }\bibfield  {title} {\bibinfo {title} {Advanced
  capabilities for materials modelling with q uantum espresso},\ }\href
  {http://stacks.iop.org/0953-8984/29/i=46/a=465901} {\bibfield  {journal}
  {\bibinfo  {journal} {Journal of Physics: Condensed Matter}\ }\textbf
  {\bibinfo {volume} {29}},\ \bibinfo {pages} {465901} (\bibinfo {year}
  {2017})}\BibitemShut {NoStop}%
\bibitem [{\citenamefont {Perdew}\ \emph {et~al.}(1996)\citenamefont {Perdew},
  \citenamefont {Burke},\ and\ \citenamefont {Ernzerhof}}]{PBE}%
  \BibitemOpen
  \bibfield  {author} {\bibinfo {author} {\bibfnamefont {J.~P.}\ \bibnamefont
  {Perdew}}, \bibinfo {author} {\bibfnamefont {K.}~\bibnamefont {Burke}},\ and\
  \bibinfo {author} {\bibfnamefont {M.}~\bibnamefont {Ernzerhof}},\ }\bibfield
  {title} {\bibinfo {title} {Generalized gradient approximation made simple},\
  }\href {https://doi.org/10.1103/PhysRevLett.77.3865} {\bibfield  {journal}
  {\bibinfo  {journal} {Phys. Rev. Lett.}\ }\textbf {\bibinfo {volume} {77}},\
  \bibinfo {pages} {3865} (\bibinfo {year} {1996})}\BibitemShut {NoStop}%
\bibitem [{\citenamefont {Kresse}\ and\ \citenamefont {Joubert}(1999)}]{PAW}%
  \BibitemOpen
  \bibfield  {author} {\bibinfo {author} {\bibfnamefont {G.}~\bibnamefont
  {Kresse}}\ and\ \bibinfo {author} {\bibfnamefont {D.}~\bibnamefont
  {Joubert}},\ }\bibfield  {title} {\bibinfo {title} {From ultrasoft
  pseudopotentials to the projector augmented-wave method},\ }\href
  {https://doi.org/10.1103/PhysRevB.59.1758} {\bibfield  {journal} {\bibinfo
  {journal} {Phys. Rev. B}\ }\textbf {\bibinfo {volume} {59}},\ \bibinfo
  {pages} {1758} (\bibinfo {year} {1999})}\BibitemShut {NoStop}%
\bibitem [{\citenamefont {{Dal Corso}}(2014)}]{pslibrary}%
  \BibitemOpen
  \bibfield  {author} {\bibinfo {author} {\bibfnamefont {A.}~\bibnamefont {{Dal
  Corso}}},\ }\bibfield  {title} {\bibinfo {title} {Pseudopotentials periodic
  table: From h to pu},\ }\href
  {https://doi.org/https://doi.org/10.1016/j.commatsci.2014.07.043} {\bibfield
  {journal} {\bibinfo  {journal} {Computational Materials Science}\ }\textbf
  {\bibinfo {volume} {95}},\ \bibinfo {pages} {337} (\bibinfo {year}
  {2014})}\BibitemShut {NoStop}%
\bibitem [{\citenamefont {Brumme}\ \emph {et~al.}(2015)\citenamefont {Brumme},
  \citenamefont {Calandra},\ and\ \citenamefont {Mauri}}]{Efield}%
  \BibitemOpen
  \bibfield  {author} {\bibinfo {author} {\bibfnamefont {T.}~\bibnamefont
  {Brumme}}, \bibinfo {author} {\bibfnamefont {M.}~\bibnamefont {Calandra}},\
  and\ \bibinfo {author} {\bibfnamefont {F.}~\bibnamefont {Mauri}},\ }\bibfield
   {title} {\bibinfo {title} {First-principles theory of field-effect doping in
  transition-metal dichalcogenides: Structural properties, electronic
  structure, hall coefficient, and electrical conductivity},\ }\href
  {https://doi.org/10.1103/PhysRevB.91.155436} {\bibfield  {journal} {\bibinfo
  {journal} {Phys. Rev. B}\ }\textbf {\bibinfo {volume} {91}},\ \bibinfo
  {pages} {155436} (\bibinfo {year} {2015})}\BibitemShut {NoStop}%
\bibitem [{\citenamefont {Agapito}\ \emph {et~al.}(2013)\citenamefont
  {Agapito}, \citenamefont {Ferretti}, \citenamefont {Calzolari}, \citenamefont
  {Curtarolo},\ and\ \citenamefont {Buongiorno~Nardelli}}]{PAO1}%
  \BibitemOpen
  \bibfield  {author} {\bibinfo {author} {\bibfnamefont {L.~A.}\ \bibnamefont
  {Agapito}}, \bibinfo {author} {\bibfnamefont {A.}~\bibnamefont {Ferretti}},
  \bibinfo {author} {\bibfnamefont {A.}~\bibnamefont {Calzolari}}, \bibinfo
  {author} {\bibfnamefont {S.}~\bibnamefont {Curtarolo}},\ and\ \bibinfo
  {author} {\bibfnamefont {M.}~\bibnamefont {Buongiorno~Nardelli}},\ }\bibfield
   {title} {\bibinfo {title} {Effective and accurate representation of extended
  bloch states on finite hilbert spaces},\ }\href
  {https://doi.org/10.1103/PhysRevB.88.165127} {\bibfield  {journal} {\bibinfo
  {journal} {Phys. Rev. B}\ }\textbf {\bibinfo {volume} {88}},\ \bibinfo
  {pages} {165127} (\bibinfo {year} {2013})}\BibitemShut {NoStop}%
\bibitem [{\citenamefont {Agapito}\ \emph {et~al.}(2015)\citenamefont
  {Agapito}, \citenamefont {Curtarolo},\ and\ \citenamefont
  {Buongiorno~Nardelli}}]{PAO2}%
  \BibitemOpen
  \bibfield  {author} {\bibinfo {author} {\bibfnamefont {L.~A.}\ \bibnamefont
  {Agapito}}, \bibinfo {author} {\bibfnamefont {S.}~\bibnamefont {Curtarolo}},\
  and\ \bibinfo {author} {\bibfnamefont {M.}~\bibnamefont
  {Buongiorno~Nardelli}},\ }\bibfield  {title} {\bibinfo {title} {Reformulation
  of $\mathrm{DFT}+u$ as a pseudohybrid hubbard density functional for
  accelerated materials discovery},\ }\href
  {https://doi.org/10.1103/PhysRevX.5.011006} {\bibfield  {journal} {\bibinfo
  {journal} {Phys. Rev. X}\ }\textbf {\bibinfo {volume} {5}},\ \bibinfo {pages}
  {011006} (\bibinfo {year} {2015})}\BibitemShut {NoStop}%
\bibitem [{\citenamefont {Agapito}\ \emph
  {et~al.}(2016{\natexlab{a}})\citenamefont {Agapito}, \citenamefont {Fornari},
  \citenamefont {Ceresoli}, \citenamefont {Ferretti}, \citenamefont
  {Curtarolo},\ and\ \citenamefont {Buongiorno~Nardelli}}]{PAO3}%
  \BibitemOpen
  \bibfield  {author} {\bibinfo {author} {\bibfnamefont {L.~A.}\ \bibnamefont
  {Agapito}}, \bibinfo {author} {\bibfnamefont {M.}~\bibnamefont {Fornari}},
  \bibinfo {author} {\bibfnamefont {D.}~\bibnamefont {Ceresoli}}, \bibinfo
  {author} {\bibfnamefont {A.}~\bibnamefont {Ferretti}}, \bibinfo {author}
  {\bibfnamefont {S.}~\bibnamefont {Curtarolo}},\ and\ \bibinfo {author}
  {\bibfnamefont {M.}~\bibnamefont {Buongiorno~Nardelli}},\ }\bibfield  {title}
  {\bibinfo {title} {Accurate tight-binding hamiltonians for two-dimensional
  and layered materials},\ }\href {https://doi.org/10.1103/PhysRevB.93.125137}
  {\bibfield  {journal} {\bibinfo  {journal} {Phys. Rev. B}\ }\textbf {\bibinfo
  {volume} {93}},\ \bibinfo {pages} {125137} (\bibinfo {year}
  {2016}{\natexlab{a}})}\BibitemShut {NoStop}%
\bibitem [{\citenamefont {Agapito}\ \emph
  {et~al.}(2016{\natexlab{b}})\citenamefont {Agapito}, \citenamefont
  {Ismail-Beigi}, \citenamefont {Curtarolo}, \citenamefont {Fornari},\ and\
  \citenamefont {Buongiorno~Nardelli}}]{PAO4}%
  \BibitemOpen
  \bibfield  {author} {\bibinfo {author} {\bibfnamefont {L.~A.}\ \bibnamefont
  {Agapito}}, \bibinfo {author} {\bibfnamefont {S.}~\bibnamefont
  {Ismail-Beigi}}, \bibinfo {author} {\bibfnamefont {S.}~\bibnamefont
  {Curtarolo}}, \bibinfo {author} {\bibfnamefont {M.}~\bibnamefont {Fornari}},\
  and\ \bibinfo {author} {\bibfnamefont {M.}~\bibnamefont
  {Buongiorno~Nardelli}},\ }\bibfield  {title} {\bibinfo {title} {Accurate
  tight-binding hamiltonian matrices from ab initio calculations: Minimal basis
  sets},\ }\href {https://doi.org/10.1103/PhysRevB.93.035104} {\bibfield
  {journal} {\bibinfo  {journal} {Phys. Rev. B}\ }\textbf {\bibinfo {volume}
  {93}},\ \bibinfo {pages} {035104} (\bibinfo {year}
  {2016}{\natexlab{b}})}\BibitemShut {NoStop}%
\bibitem [{\citenamefont {Buongiorno~Nardelli}\ \emph
  {et~al.}(2018)\citenamefont {Buongiorno~Nardelli}, \citenamefont {Cerasoli},
  \citenamefont {Costa}, \citenamefont {Curtarolo}, \citenamefont {Gennaro},
  \citenamefont {Fornari}, \citenamefont {Liyanage}, \citenamefont {Supka},\
  and\ \citenamefont {Wang}}]{PAO5}%
  \BibitemOpen
  \bibfield  {author} {\bibinfo {author} {\bibfnamefont {M.}~\bibnamefont
  {Buongiorno~Nardelli}}, \bibinfo {author} {\bibfnamefont {F.~T.}\
  \bibnamefont {Cerasoli}}, \bibinfo {author} {\bibfnamefont {M.}~\bibnamefont
  {Costa}}, \bibinfo {author} {\bibfnamefont {S.}~\bibnamefont {Curtarolo}},
  \bibinfo {author} {\bibfnamefont {R.~D.}\ \bibnamefont {Gennaro}}, \bibinfo
  {author} {\bibfnamefont {M.}~\bibnamefont {Fornari}}, \bibinfo {author}
  {\bibfnamefont {L.}~\bibnamefont {Liyanage}}, \bibinfo {author}
  {\bibfnamefont {A.~R.}\ \bibnamefont {Supka}},\ and\ \bibinfo {author}
  {\bibfnamefont {H.}~\bibnamefont {Wang}},\ }\bibfield  {title} {\bibinfo
  {title} {Paoflow: A utility to construct and operate on ab initio
  hamiltonians from the projections of electronic wavefunctions on atomic
  orbital bases, including characterization of topological materials},\ }\href
  {https://doi.org/https://doi.org/10.1016/j.commatsci.2017.11.034} {\bibfield
  {journal} {\bibinfo  {journal} {Computational Materials Science}\ }\textbf
  {\bibinfo {volume} {143}},\ \bibinfo {pages} {462 } (\bibinfo {year}
  {2018})}\BibitemShut {NoStop}%
\bibitem [{\citenamefont {Cerasoli}\ \emph {et~al.}(2021)\citenamefont
  {Cerasoli}, \citenamefont {Supka}, \citenamefont {Jayaraj}, \citenamefont
  {Costa}, \citenamefont {Siloi}, \citenamefont {Sławińska}, \citenamefont
  {Curtarolo}, \citenamefont {Fornari}, \citenamefont {Ceresoli},\ and\
  \citenamefont {{Buongiorno Nardelli}}}]{PAO6}%
  \BibitemOpen
  \bibfield  {author} {\bibinfo {author} {\bibfnamefont {F.~T.}\ \bibnamefont
  {Cerasoli}}, \bibinfo {author} {\bibfnamefont {A.~R.}\ \bibnamefont {Supka}},
  \bibinfo {author} {\bibfnamefont {A.}~\bibnamefont {Jayaraj}}, \bibinfo
  {author} {\bibfnamefont {M.}~\bibnamefont {Costa}}, \bibinfo {author}
  {\bibfnamefont {I.}~\bibnamefont {Siloi}}, \bibinfo {author} {\bibfnamefont
  {J.}~\bibnamefont {Sławińska}}, \bibinfo {author} {\bibfnamefont
  {S.}~\bibnamefont {Curtarolo}}, \bibinfo {author} {\bibfnamefont
  {M.}~\bibnamefont {Fornari}}, \bibinfo {author} {\bibfnamefont
  {D.}~\bibnamefont {Ceresoli}},\ and\ \bibinfo {author} {\bibfnamefont
  {M.}~\bibnamefont {{Buongiorno Nardelli}}},\ }\bibfield  {title} {\bibinfo
  {title} {Advanced modeling of materials with paoflow 2.0: New features and
  software design},\ }\href
  {https://doi.org/https://doi.org/10.1016/j.commatsci.2021.110828} {\bibfield
  {journal} {\bibinfo  {journal} {Computational Materials Science}\ }\textbf
  {\bibinfo {volume} {200}},\ \bibinfo {pages} {110828} (\bibinfo {year}
  {2021})}\BibitemShut {NoStop}%
\bibitem [{\citenamefont {Menezes}\ and\ \citenamefont
  {Capaz}(2018)}]{MENEZES2018411}%
  \BibitemOpen
  \bibfield  {author} {\bibinfo {author} {\bibfnamefont {M.~G.}\ \bibnamefont
  {Menezes}}\ and\ \bibinfo {author} {\bibfnamefont {R.~B.}\ \bibnamefont
  {Capaz}},\ }\bibfield  {title} {\bibinfo {title} {Tight binding
  parametrization of few-layer black phosphorus from first-principles
  calculations},\ }\href
  {https://doi.org/https://doi.org/10.1016/j.commatsci.2017.11.039} {\bibfield
  {journal} {\bibinfo  {journal} {Computational Materials Science}\ }\textbf
  {\bibinfo {volume} {143}},\ \bibinfo {pages} {411} (\bibinfo {year}
  {2018})}\BibitemShut {NoStop}%
\bibitem [{\citenamefont {Costa}\ \emph
  {et~al.}(2018{\natexlab{a}})\citenamefont {Costa}, \citenamefont {Nardelli},
  \citenamefont {Fazzio},\ and\ \citenamefont {Costa}}]{adatoms}%
  \BibitemOpen
  \bibfield  {author} {\bibinfo {author} {\bibfnamefont {M.}~\bibnamefont
  {Costa}}, \bibinfo {author} {\bibfnamefont {M.~B.}\ \bibnamefont {Nardelli}},
  \bibinfo {author} {\bibfnamefont {A.}~\bibnamefont {Fazzio}},\ and\ \bibinfo
  {author} {\bibfnamefont {A.~T.}\ \bibnamefont {Costa}},\ }\href
  {https://doi.org/10.48550/ARXIV.1808.00347} {\bibinfo {title} {Long range
  dynamical coupling between magnetic adatoms mediated by a 2d topological
  insulator}} (\bibinfo {year} {2018}{\natexlab{a}})\BibitemShut {NoStop}%
\bibitem [{\citenamefont {Costa}\ \emph {et~al.}(2020)\citenamefont {Costa},
  \citenamefont {Peres}, \citenamefont {Fern\'andez-Rossier},\ and\
  \citenamefont {Costa}}]{fegete}%
  \BibitemOpen
  \bibfield  {author} {\bibinfo {author} {\bibfnamefont {M.}~\bibnamefont
  {Costa}}, \bibinfo {author} {\bibfnamefont {N.~M.~R.}\ \bibnamefont {Peres}},
  \bibinfo {author} {\bibfnamefont {J.}~\bibnamefont {Fern\'andez-Rossier}},\
  and\ \bibinfo {author} {\bibfnamefont {A.~T.}\ \bibnamefont {Costa}},\
  }\bibfield  {title} {\bibinfo {title} {Nonreciprocal magnons in a
  two-dimensional crystal with out-of-plane magnetization},\ }\href
  {https://doi.org/10.1103/PhysRevB.102.014450} {\bibfield  {journal} {\bibinfo
   {journal} {Phys. Rev. B}\ }\textbf {\bibinfo {volume} {102}},\ \bibinfo
  {pages} {014450} (\bibinfo {year} {2020})}\BibitemShut {NoStop}%
\bibitem [{\citenamefont {Costa}\ \emph {et~al.}(2021)\citenamefont {Costa},
  \citenamefont {Schleder}, \citenamefont {Acosta}, \citenamefont {Padilha},
  \citenamefont {Cerasoli}, \citenamefont {Nardelli},\ and\ \citenamefont
  {Fazzio}}]{hoti}%
  \BibitemOpen
  \bibfield  {author} {\bibinfo {author} {\bibfnamefont {M.}~\bibnamefont
  {Costa}}, \bibinfo {author} {\bibfnamefont {G.~R.}\ \bibnamefont {Schleder}},
  \bibinfo {author} {\bibfnamefont {C.~M.}\ \bibnamefont {Acosta}}, \bibinfo
  {author} {\bibfnamefont {A.~C.~M.}\ \bibnamefont {Padilha}}, \bibinfo
  {author} {\bibfnamefont {F.}~\bibnamefont {Cerasoli}}, \bibinfo {author}
  {\bibfnamefont {M.~B.}\ \bibnamefont {Nardelli}},\ and\ \bibinfo {author}
  {\bibfnamefont {A.}~\bibnamefont {Fazzio}},\ }\bibfield  {title} {\bibinfo
  {title} {Discovery of higher-order topological insulators using the spin hall
  conductivity as a topology signature},\ }\href
  {https://doi.org/10.1038/s41524-021-00518-4} {\bibfield  {journal} {\bibinfo
  {journal} {npj Computational Materials}\ }\textbf {\bibinfo {volume} {7}},\
  \bibinfo {pages} {49} (\bibinfo {year} {2021})}\BibitemShut {NoStop}%
\bibitem [{\citenamefont {Heath}\ \emph {et~al.}(2020)\citenamefont {Heath},
  \citenamefont {Costa}, \citenamefont {Buongiorno-Nardelli},\ and\
  \citenamefont {Kuroda}}]{cri3-graphene}%
  \BibitemOpen
  \bibfield  {author} {\bibinfo {author} {\bibfnamefont {J.~J.}\ \bibnamefont
  {Heath}}, \bibinfo {author} {\bibfnamefont {M.}~\bibnamefont {Costa}},
  \bibinfo {author} {\bibfnamefont {M.}~\bibnamefont {Buongiorno-Nardelli}},\
  and\ \bibinfo {author} {\bibfnamefont {M.~A.}\ \bibnamefont {Kuroda}},\
  }\bibfield  {title} {\bibinfo {title} {Role of quantum confinement and
  interlayer coupling in ${\mathrm{cri}}_{3}$-graphene magnetic tunnel
  junctions},\ }\href {https://doi.org/10.1103/PhysRevB.101.195439} {\bibfield
  {journal} {\bibinfo  {journal} {Phys. Rev. B}\ }\textbf {\bibinfo {volume}
  {101}},\ \bibinfo {pages} {195439} (\bibinfo {year} {2020})}\BibitemShut
  {NoStop}%
\bibitem [{\citenamefont {Costa}\ \emph {et~al.}(2019)\citenamefont {Costa},
  \citenamefont {Schleder}, \citenamefont {Buongiorno~Nardelli}, \citenamefont
  {Lewenkopf},\ and\ \citenamefont {Fazzio}}]{Costa2019}%
  \BibitemOpen
  \bibfield  {author} {\bibinfo {author} {\bibfnamefont {M.}~\bibnamefont
  {Costa}}, \bibinfo {author} {\bibfnamefont {G.~R.}\ \bibnamefont {Schleder}},
  \bibinfo {author} {\bibfnamefont {M.}~\bibnamefont {Buongiorno~Nardelli}},
  \bibinfo {author} {\bibfnamefont {C.}~\bibnamefont {Lewenkopf}},\ and\
  \bibinfo {author} {\bibfnamefont {A.}~\bibnamefont {Fazzio}},\ }\bibfield
  {title} {\bibinfo {title} {Toward realistic amorphous topological
  insulators},\ }\href {https://doi.org/10.1021/acs.nanolett.9b03881}
  {\bibfield  {journal} {\bibinfo  {journal} {Nano Letters}\ }\textbf {\bibinfo
  {volume} {19}},\ \bibinfo {pages} {8941} (\bibinfo {year}
  {2019})}\BibitemShut {NoStop}%
\bibitem [{\citenamefont {Costa}\ \emph
  {et~al.}(2018{\natexlab{b}})\citenamefont {Costa}, \citenamefont {Costa},
  \citenamefont {Freitas}, \citenamefont {Schmidt}, \citenamefont
  {Buongiorno~Nardelli},\ and\ \citenamefont {Fazzio}}]{Costa2018}%
  \BibitemOpen
  \bibfield  {author} {\bibinfo {author} {\bibfnamefont {M.}~\bibnamefont
  {Costa}}, \bibinfo {author} {\bibfnamefont {A.~T.}\ \bibnamefont {Costa}},
  \bibinfo {author} {\bibfnamefont {W.~A.}\ \bibnamefont {Freitas}}, \bibinfo
  {author} {\bibfnamefont {T.~M.}\ \bibnamefont {Schmidt}}, \bibinfo {author}
  {\bibfnamefont {M.}~\bibnamefont {Buongiorno~Nardelli}},\ and\ \bibinfo
  {author} {\bibfnamefont {A.}~\bibnamefont {Fazzio}},\ }\bibfield  {title}
  {\bibinfo {title} {Controlling topological states in topological/normal
  insulator heterostructures},\ }\href
  {https://doi.org/10.1021/acsomega.8b01836} {\bibfield  {journal} {\bibinfo
  {journal} {ACS Omega}\ }\textbf {\bibinfo {volume} {3}},\ \bibinfo {pages}
  {15900} (\bibinfo {year} {2018}{\natexlab{b}})}\BibitemShut {NoStop}%
\bibitem [{\citenamefont {Liu}\ and\ \citenamefont
  {Culcer}(2023)}]{liu2023dominance}%
  \BibitemOpen
  \bibfield  {author} {\bibinfo {author} {\bibfnamefont {H.}~\bibnamefont
  {Liu}}\ and\ \bibinfo {author} {\bibfnamefont {D.}~\bibnamefont {Culcer}},\
  }\href@noop {} {\bibinfo {title} {Dominance of extrinsic scattering
  mechanisms in the orbital hall effect: graphene, transition metal
  dichalcogenides and topological antiferromagnets}} (\bibinfo {year} {2023}),\
  \Eprint {https://arxiv.org/abs/2308.14878} {arXiv:2308.14878
  [cond-mat.mes-hall]} \BibitemShut {NoStop}%
\bibitem [{\citenamefont
  {Dimitrova}(2005)}]{DimitrovaVertex-PhysRevB.71.245327}%
  \BibitemOpen
  \bibfield  {author} {\bibinfo {author} {\bibfnamefont {O.~V.}\ \bibnamefont
  {Dimitrova}},\ }\bibfield  {title} {\bibinfo {title} {Spin-hall conductivity
  in a two-dimensional rashba electron gas},\ }\href
  {https://doi.org/10.1103/PhysRevB.71.245327} {\bibfield  {journal} {\bibinfo
  {journal} {Phys. Rev. B}\ }\textbf {\bibinfo {volume} {71}},\ \bibinfo
  {pages} {245327} (\bibinfo {year} {2005})}\BibitemShut {NoStop}%
\bibitem [{\citenamefont {Hitomi}\ \emph {et~al.}(2021)\citenamefont {Hitomi},
  \citenamefont {Kawakami},\ and\ \citenamefont
  {Koshino}}]{HOTI-Phosphorene-PhysRevB.104.125302}%
  \BibitemOpen
  \bibfield  {author} {\bibinfo {author} {\bibfnamefont {M.}~\bibnamefont
  {Hitomi}}, \bibinfo {author} {\bibfnamefont {T.}~\bibnamefont {Kawakami}},\
  and\ \bibinfo {author} {\bibfnamefont {M.}~\bibnamefont {Koshino}},\
  }\bibfield  {title} {\bibinfo {title} {Multiorbital edge and corner states in
  black phosphorene},\ }\href {https://doi.org/10.1103/PhysRevB.104.125302}
  {\bibfield  {journal} {\bibinfo  {journal} {Phys. Rev. B}\ }\textbf {\bibinfo
  {volume} {104}},\ \bibinfo {pages} {125302} (\bibinfo {year}
  {2021})}\BibitemShut {NoStop}%
\bibitem [{\citenamefont {Ezawa}(2018)}]{HOTI-Phosphorene-PhysRevB.98.045125}%
  \BibitemOpen
  \bibfield  {author} {\bibinfo {author} {\bibfnamefont {M.}~\bibnamefont
  {Ezawa}},\ }\bibfield  {title} {\bibinfo {title} {Minimal models for
  wannier-type higher-order topological insulators and phosphorene},\ }\href
  {https://doi.org/10.1103/PhysRevB.98.045125} {\bibfield  {journal} {\bibinfo
  {journal} {Phys. Rev. B}\ }\textbf {\bibinfo {volume} {98}},\ \bibinfo
  {pages} {045125} (\bibinfo {year} {2018})}\BibitemShut {NoStop}%
\bibitem [{\citenamefont {Lee}\ \emph {et~al.}(2022)\citenamefont {Lee},
  \citenamefont {Choi},\ and\ \citenamefont
  {Lee}}]{H-Woo-Symmetry_PhysRevB.105.035142}%
  \BibitemOpen
  \bibfield  {author} {\bibinfo {author} {\bibfnamefont {H.}~\bibnamefont
  {Lee}}, \bibinfo {author} {\bibfnamefont {B.}~\bibnamefont {Choi}},\ and\
  \bibinfo {author} {\bibfnamefont {H.-W.}\ \bibnamefont {Lee}},\ }\bibfield
  {title} {\bibinfo {title} {Orientational dependence of intrinsic orbital and
  spin hall effects in hcp structure materials},\ }\href
  {https://doi.org/10.1103/PhysRevB.105.035142} {\bibfield  {journal} {\bibinfo
   {journal} {Phys. Rev. B}\ }\textbf {\bibinfo {volume} {105}},\ \bibinfo
  {pages} {035142} (\bibinfo {year} {2022})}\BibitemShut {NoStop}%
\bibitem [{\citenamefont {Jungwirth}\ \emph {et~al.}(2012)\citenamefont
  {Jungwirth}, \citenamefont {Wunderlich},\ and\ \citenamefont
  {Olejn{\'{\i}}k}}]{SHE-Devices-Jungwirth2012}%
  \BibitemOpen
  \bibfield  {author} {\bibinfo {author} {\bibfnamefont {T.}~\bibnamefont
  {Jungwirth}}, \bibinfo {author} {\bibfnamefont {J.}~\bibnamefont
  {Wunderlich}},\ and\ \bibinfo {author} {\bibfnamefont {K.}~\bibnamefont
  {Olejn{\'{\i}}k}},\ }\bibfield  {title} {\bibinfo {title} {Spin hall effect
  devices},\ }\href {https://doi.org/10.1038/nmat3279} {\bibfield  {journal}
  {\bibinfo  {journal} {Nature Materials}\ }\textbf {\bibinfo {volume} {11}},\
  \bibinfo {pages} {382} (\bibinfo {year} {2012})}\BibitemShut {NoStop}%
\bibitem [{\citenamefont {Marui}\ \emph {et~al.}(2023)\citenamefont {Marui},
  \citenamefont {Kawaguchi}, \citenamefont {Sumi}, \citenamefont {Awano},
  \citenamefont {Nakamura},\ and\ \citenamefont {Hayashi}}]{marui2023spin}%
  \BibitemOpen
  \bibfield  {author} {\bibinfo {author} {\bibfnamefont {Y.}~\bibnamefont
  {Marui}}, \bibinfo {author} {\bibfnamefont {M.}~\bibnamefont {Kawaguchi}},
  \bibinfo {author} {\bibfnamefont {S.}~\bibnamefont {Sumi}}, \bibinfo {author}
  {\bibfnamefont {H.}~\bibnamefont {Awano}}, \bibinfo {author} {\bibfnamefont
  {K.}~\bibnamefont {Nakamura}},\ and\ \bibinfo {author} {\bibfnamefont
  {M.}~\bibnamefont {Hayashi}},\ }\href@noop {} {\bibinfo {title} {Spin and
  orbital hall currents detected via current induced magneto-optical kerr
  effect in v and pt}} (\bibinfo {year} {2023}),\ \Eprint
  {https://arxiv.org/abs/2306.09585} {arXiv:2306.09585 [cond-mat.mes-hall]}
  \BibitemShut {NoStop}%
\bibitem [{\citenamefont {Kumar}\ and\ \citenamefont
  {Kumar}(2023)}]{kumar2023ultrafast}%
  \BibitemOpen
  \bibfield  {author} {\bibinfo {author} {\bibfnamefont {S.}~\bibnamefont
  {Kumar}}\ and\ \bibinfo {author} {\bibfnamefont {S.}~\bibnamefont {Kumar}},\
  }\href@noop {} {\bibinfo {title} {Ultrafast thz probing of nonlocal orbital
  current in transverse multilayer metallic heterostructures}} (\bibinfo {year}
  {2023}),\ \Eprint {https://arxiv.org/abs/2306.17027} {arXiv:2306.17027
  [cond-mat.mtrl-sci]} \BibitemShut {NoStop}%
\bibitem [{\citenamefont {Lyalin}\ \emph {et~al.}(2023)\citenamefont {Lyalin},
  \citenamefont {Alikhah}, \citenamefont {Berritta}, \citenamefont {Oppeneer},\
  and\ \citenamefont {Kawakami}}]{lyalin2023magnetooptical}%
  \BibitemOpen
  \bibfield  {author} {\bibinfo {author} {\bibfnamefont {I.}~\bibnamefont
  {Lyalin}}, \bibinfo {author} {\bibfnamefont {S.}~\bibnamefont {Alikhah}},
  \bibinfo {author} {\bibfnamefont {M.}~\bibnamefont {Berritta}}, \bibinfo
  {author} {\bibfnamefont {P.~M.}\ \bibnamefont {Oppeneer}},\ and\ \bibinfo
  {author} {\bibfnamefont {R.~K.}\ \bibnamefont {Kawakami}},\ }\href@noop {}
  {\bibinfo {title} {Magneto-optical detection of the orbital hall effect in
  chromium}} (\bibinfo {year} {2023}),\ \Eprint
  {https://arxiv.org/abs/2306.10673} {arXiv:2306.10673 [cond-mat.mes-hall]}
  \BibitemShut {NoStop}%
\bibitem [{\citenamefont {Cysne}\ \emph {et~al.}(2023)\citenamefont {Cysne},
  \citenamefont {Guimar\~aes}, \citenamefont {Canonico}, \citenamefont {Costa},
  \citenamefont {Rappoport},\ and\ \citenamefont
  {Muniz}}]{Cysne-PhysRevB.107.115402}%
  \BibitemOpen
  \bibfield  {author} {\bibinfo {author} {\bibfnamefont {T.~P.}\ \bibnamefont
  {Cysne}}, \bibinfo {author} {\bibfnamefont {F.~S.~M.}\ \bibnamefont
  {Guimar\~aes}}, \bibinfo {author} {\bibfnamefont {L.~M.}\ \bibnamefont
  {Canonico}}, \bibinfo {author} {\bibfnamefont {M.}~\bibnamefont {Costa}},
  \bibinfo {author} {\bibfnamefont {T.~G.}\ \bibnamefont {Rappoport}},\ and\
  \bibinfo {author} {\bibfnamefont {R.~B.}\ \bibnamefont {Muniz}},\ }\bibfield
  {title} {\bibinfo {title} {Orbital magnetoelectric effect in nanoribbons of
  transition metal dichalcogenides},\ }\href
  {https://doi.org/10.1103/PhysRevB.107.115402} {\bibfield  {journal} {\bibinfo
   {journal} {Phys. Rev. B}\ }\textbf {\bibinfo {volume} {107}},\ \bibinfo
  {pages} {115402} (\bibinfo {year} {2023})}\BibitemShut {NoStop}%
\bibitem [{\citenamefont {Ribeiro-Soares}\ \emph {et~al.}(2015)\citenamefont
  {Ribeiro-Soares}, \citenamefont {Almeida}, \citenamefont
  {Can\ifmmode~\mbox{\c{c}}\else \c{c}\fi{}ado}, \citenamefont {Dresselhaus},\
  and\ \citenamefont {Jorio}}]{GroupTheory-PhysRevB.91.205421}%
  \BibitemOpen
  \bibfield  {author} {\bibinfo {author} {\bibfnamefont {J.}~\bibnamefont
  {Ribeiro-Soares}}, \bibinfo {author} {\bibfnamefont {R.~M.}\ \bibnamefont
  {Almeida}}, \bibinfo {author} {\bibfnamefont {L.~G.}\ \bibnamefont
  {Can\ifmmode~\mbox{\c{c}}\else \c{c}\fi{}ado}}, \bibinfo {author}
  {\bibfnamefont {M.~S.}\ \bibnamefont {Dresselhaus}},\ and\ \bibinfo {author}
  {\bibfnamefont {A.}~\bibnamefont {Jorio}},\ }\bibfield  {title} {\bibinfo
  {title} {Group theory for structural analysis and lattice vibrations in
  phosphorene systems},\ }\href {https://doi.org/10.1103/PhysRevB.91.205421}
  {\bibfield  {journal} {\bibinfo  {journal} {Phys. Rev. B}\ }\textbf {\bibinfo
  {volume} {91}},\ \bibinfo {pages} {205421} (\bibinfo {year}
  {2015})}\BibitemShut {NoStop}%
\bibitem [{\citenamefont {Hu}\ and\ \citenamefont
  {Dong}(2015)}]{Structural-Phase-PhysRevB.92.064114}%
  \BibitemOpen
  \bibfield  {author} {\bibinfo {author} {\bibfnamefont {T.}~\bibnamefont
  {Hu}}\ and\ \bibinfo {author} {\bibfnamefont {J.}~\bibnamefont {Dong}},\
  }\bibfield  {title} {\bibinfo {title} {Structural phase transitions of
  phosphorene induced by applied strains},\ }\href
  {https://doi.org/10.1103/PhysRevB.92.064114} {\bibfield  {journal} {\bibinfo
  {journal} {Phys. Rev. B}\ }\textbf {\bibinfo {volume} {92}},\ \bibinfo
  {pages} {064114} (\bibinfo {year} {2015})}\BibitemShut {NoStop}%
\bibitem [{\citenamefont {Peng}\ \emph {et~al.}(2014)\citenamefont {Peng},
  \citenamefont {Wei},\ and\ \citenamefont
  {Copple}}]{Peng-Wei-Copple-PhysRevB.90.085402}%
  \BibitemOpen
  \bibfield  {author} {\bibinfo {author} {\bibfnamefont {X.}~\bibnamefont
  {Peng}}, \bibinfo {author} {\bibfnamefont {Q.}~\bibnamefont {Wei}},\ and\
  \bibinfo {author} {\bibfnamefont {A.}~\bibnamefont {Copple}},\ }\bibfield
  {title} {\bibinfo {title} {Strain-engineered direct-indirect band gap
  transition and its mechanism in two-dimensional phosphorene},\ }\href
  {https://doi.org/10.1103/PhysRevB.90.085402} {\bibfield  {journal} {\bibinfo
  {journal} {Phys. Rev. B}\ }\textbf {\bibinfo {volume} {90}},\ \bibinfo
  {pages} {085402} (\bibinfo {year} {2014})}\BibitemShut {NoStop}%
\bibitem [{\citenamefont {Midtvedt}\ \emph {et~al.}(2016)\citenamefont
  {Midtvedt}, \citenamefont {Lewenkopf},\ and\ \citenamefont
  {Croy}}]{Lewenkopf-Midtvedt2016}%
  \BibitemOpen
  \bibfield  {author} {\bibinfo {author} {\bibfnamefont {D.}~\bibnamefont
  {Midtvedt}}, \bibinfo {author} {\bibfnamefont {C.~H.}\ \bibnamefont
  {Lewenkopf}},\ and\ \bibinfo {author} {\bibfnamefont {A.}~\bibnamefont
  {Croy}},\ }\bibfield  {title} {\bibinfo {title}
  {Strain{\textendash}displacement relations for strain engineering in
  single-layer 2d materials},\ }\href
  {https://doi.org/10.1088/2053-1583/3/1/011005} {\bibfield  {journal}
  {\bibinfo  {journal} {2D Materials}\ }\textbf {\bibinfo {volume} {3}},\
  \bibinfo {pages} {011005} (\bibinfo {year} {2016})}\BibitemShut {NoStop}%
\bibitem [{\citenamefont {Seemann}\ \emph {et~al.}(2015)\citenamefont
  {Seemann}, \citenamefont {K\"odderitzsch}, \citenamefont {Wimmer},\ and\
  \citenamefont {Ebert}}]{Ebert-Symmetry-tensor}%
  \BibitemOpen
  \bibfield  {author} {\bibinfo {author} {\bibfnamefont {M.}~\bibnamefont
  {Seemann}}, \bibinfo {author} {\bibfnamefont {D.}~\bibnamefont
  {K\"odderitzsch}}, \bibinfo {author} {\bibfnamefont {S.}~\bibnamefont
  {Wimmer}},\ and\ \bibinfo {author} {\bibfnamefont {H.}~\bibnamefont
  {Ebert}},\ }\bibfield  {title} {\bibinfo {title} {Symmetry-imposed shape of
  linear response tensors},\ }\href
  {https://doi.org/10.1103/PhysRevB.92.155138} {\bibfield  {journal} {\bibinfo
  {journal} {Phys. Rev. B}\ }\textbf {\bibinfo {volume} {92}},\ \bibinfo
  {pages} {155138} (\bibinfo {year} {2015})}\BibitemShut {NoStop}%
\bibitem [{\citenamefont {Roy}\ \emph {et~al.}(2022)\citenamefont {Roy},
  \citenamefont {Guimar\~aes},\ and\ \citenamefont
  {S\l{}awi\ifmmode~\acute{n}\else \'{n}\fi{}ska}}]{Marcosguimaraes}%
  \BibitemOpen
  \bibfield  {author} {\bibinfo {author} {\bibfnamefont {A.}~\bibnamefont
  {Roy}}, \bibinfo {author} {\bibfnamefont {M.~H.~D.}\ \bibnamefont
  {Guimar\~aes}},\ and\ \bibinfo {author} {\bibfnamefont {J.}~\bibnamefont
  {S\l{}awi\ifmmode~\acute{n}\else \'{n}\fi{}ska}},\ }\bibfield  {title}
  {\bibinfo {title} {Unconventional spin hall effects in nonmagnetic solids},\
  }\href {https://doi.org/10.1103/PhysRevMaterials.6.045004} {\bibfield
  {journal} {\bibinfo  {journal} {Phys. Rev. Mater.}\ }\textbf {\bibinfo
  {volume} {6}},\ \bibinfo {pages} {045004} (\bibinfo {year}
  {2022})}\BibitemShut {NoStop}%
\bibitem [{\citenamefont {Furukawa}\ \emph {et~al.}(2021)\citenamefont
  {Furukawa}, \citenamefont {Watanabe}, \citenamefont {Ogasawara},
  \citenamefont {Kobayashi},\ and\ \citenamefont
  {Itou}}]{C2v-Magnetoelectric-PhysRevResearch.3.023111}%
  \BibitemOpen
  \bibfield  {author} {\bibinfo {author} {\bibfnamefont {T.}~\bibnamefont
  {Furukawa}}, \bibinfo {author} {\bibfnamefont {Y.}~\bibnamefont {Watanabe}},
  \bibinfo {author} {\bibfnamefont {N.}~\bibnamefont {Ogasawara}}, \bibinfo
  {author} {\bibfnamefont {K.}~\bibnamefont {Kobayashi}},\ and\ \bibinfo
  {author} {\bibfnamefont {T.}~\bibnamefont {Itou}},\ }\bibfield  {title}
  {\bibinfo {title} {Current-induced magnetization caused by crystal chirality
  in nonmagnetic elemental tellurium},\ }\href
  {https://doi.org/10.1103/PhysRevResearch.3.023111} {\bibfield  {journal}
  {\bibinfo  {journal} {Phys. Rev. Res.}\ }\textbf {\bibinfo {volume} {3}},\
  \bibinfo {pages} {023111} (\bibinfo {year} {2021})}\BibitemShut {NoStop}%
\bibitem [{\citenamefont {Cysne}\ \emph
  {et~al.}(2021{\natexlab{b}})\citenamefont {Cysne}, \citenamefont
  {Guimar\~aes}, \citenamefont {Canonico}, \citenamefont {Rappoport},\ and\
  \citenamefont {Muniz}}]{Cysne-OMEpxpy-PhysRevB.104.165403}%
  \BibitemOpen
  \bibfield  {author} {\bibinfo {author} {\bibfnamefont {T.~P.}\ \bibnamefont
  {Cysne}}, \bibinfo {author} {\bibfnamefont {F.~S.~M.}\ \bibnamefont
  {Guimar\~aes}}, \bibinfo {author} {\bibfnamefont {L.~M.}\ \bibnamefont
  {Canonico}}, \bibinfo {author} {\bibfnamefont {T.~G.}\ \bibnamefont
  {Rappoport}},\ and\ \bibinfo {author} {\bibfnamefont {R.~B.}\ \bibnamefont
  {Muniz}},\ }\bibfield  {title} {\bibinfo {title} {Orbital magnetoelectric
  effect in zigzag nanoribbons of $p$-band systems},\ }\href
  {https://doi.org/10.1103/PhysRevB.104.165403} {\bibfield  {journal} {\bibinfo
   {journal} {Phys. Rev. B}\ }\textbf {\bibinfo {volume} {104}},\ \bibinfo
  {pages} {165403} (\bibinfo {year} {2021}{\natexlab{b}})}\BibitemShut
  {NoStop}%
\bibitem [{\citenamefont {Shinada}\ and\ \citenamefont
  {Peters}(2023)}]{Koki-Peters-PhysRevB.107.214109}%
  \BibitemOpen
  \bibfield  {author} {\bibinfo {author} {\bibfnamefont {K.}~\bibnamefont
  {Shinada}}\ and\ \bibinfo {author} {\bibfnamefont {R.}~\bibnamefont
  {Peters}},\ }\bibfield  {title} {\bibinfo {title} {Orbital
  gravitomagnetoelectric response and orbital magnetic quadrupole moment
  correction},\ }\href {https://doi.org/10.1103/PhysRevB.107.214109} {\bibfield
   {journal} {\bibinfo  {journal} {Phys. Rev. B}\ }\textbf {\bibinfo {volume}
  {107}},\ \bibinfo {pages} {214109} (\bibinfo {year} {2023})}\BibitemShut
  {NoStop}%
\bibitem [{\citenamefont {Shinada}\ \emph {et~al.}(2023)\citenamefont
  {Shinada}, \citenamefont {Kofuji},\ and\ \citenamefont
  {Peters}}]{Koki-Peters-PhysRevB.107.094106}%
  \BibitemOpen
  \bibfield  {author} {\bibinfo {author} {\bibfnamefont {K.}~\bibnamefont
  {Shinada}}, \bibinfo {author} {\bibfnamefont {A.}~\bibnamefont {Kofuji}},\
  and\ \bibinfo {author} {\bibfnamefont {R.}~\bibnamefont {Peters}},\
  }\bibfield  {title} {\bibinfo {title} {Quantum theory of the intrinsic
  orbital magnetoelectric effect in itinerant electron systems at finite
  temperatures},\ }\href {https://doi.org/10.1103/PhysRevB.107.094106}
  {\bibfield  {journal} {\bibinfo  {journal} {Phys. Rev. B}\ }\textbf {\bibinfo
  {volume} {107}},\ \bibinfo {pages} {094106} (\bibinfo {year}
  {2023})}\BibitemShut {NoStop}%
\bibitem [{\citenamefont {Hayami}\ \emph {et~al.}(2018)\citenamefont {Hayami},
  \citenamefont {Yatsushiro}, \citenamefont {Yanagi},\ and\ \citenamefont
  {Kusunose}}]{Hayami-PhysRevB.98.165110}%
  \BibitemOpen
  \bibfield  {author} {\bibinfo {author} {\bibfnamefont {S.}~\bibnamefont
  {Hayami}}, \bibinfo {author} {\bibfnamefont {M.}~\bibnamefont {Yatsushiro}},
  \bibinfo {author} {\bibfnamefont {Y.}~\bibnamefont {Yanagi}},\ and\ \bibinfo
  {author} {\bibfnamefont {H.}~\bibnamefont {Kusunose}},\ }\bibfield  {title}
  {\bibinfo {title} {Classification of atomic-scale multipoles under
  crystallographic point groups and application to linear response tensors},\
  }\href {https://doi.org/10.1103/PhysRevB.98.165110} {\bibfield  {journal}
  {\bibinfo  {journal} {Phys. Rev. B}\ }\textbf {\bibinfo {volume} {98}},\
  \bibinfo {pages} {165110} (\bibinfo {year} {2018})}\BibitemShut {NoStop}%
\bibitem [{\citenamefont {Hayami}\ \emph {et~al.}(2016)\citenamefont {Hayami},
  \citenamefont {Kusunose},\ and\ \citenamefont {Motome}}]{Hayami-JPCM-2016}%
  \BibitemOpen
  \bibfield  {author} {\bibinfo {author} {\bibfnamefont {S.}~\bibnamefont
  {Hayami}}, \bibinfo {author} {\bibfnamefont {H.}~\bibnamefont {Kusunose}},\
  and\ \bibinfo {author} {\bibfnamefont {Y.}~\bibnamefont {Motome}},\
  }\bibfield  {title} {\bibinfo {title} {Emergent spin-valley-orbital physics
  by spontaneous parity breaking},\ }\href
  {https://doi.org/10.1088/0953-8984/28/39/395601} {\bibfield  {journal}
  {\bibinfo  {journal} {Journal of Physics: Condensed Matter}\ }\textbf
  {\bibinfo {volume} {28}},\ \bibinfo {pages} {395601} (\bibinfo {year}
  {2016})}\BibitemShut {NoStop}%
\bibitem [{\citenamefont {Salemi}\ \emph {et~al.}(2019)\citenamefont {Salemi},
  \citenamefont {Berritta}, \citenamefont {Nandy},\ and\ \citenamefont
  {Oppeneer}}]{Salemi-Oppeneer-2019}%
  \BibitemOpen
  \bibfield  {author} {\bibinfo {author} {\bibfnamefont {L.}~\bibnamefont
  {Salemi}}, \bibinfo {author} {\bibfnamefont {M.}~\bibnamefont {Berritta}},
  \bibinfo {author} {\bibfnamefont {A.~K.}\ \bibnamefont {Nandy}},\ and\
  \bibinfo {author} {\bibfnamefont {P.~M.}\ \bibnamefont {Oppeneer}},\
  }\bibfield  {title} {\bibinfo {title} {Orbitally dominated rashba-edelstein
  effect in noncentrosymmetric antiferromagnets},\ }\href
  {https://doi.org/10.1038/s41467-019-13367-z} {\bibfield  {journal} {\bibinfo
  {journal} {Nature Communications}\ }\textbf {\bibinfo {volume} {10}},\
  \bibinfo {pages} {5381} (\bibinfo {year} {2019})}\BibitemShut {NoStop}%
\bibitem [{\citenamefont {Yoda}\ \emph {et~al.}(2018)\citenamefont {Yoda},
  \citenamefont {Yokoyama},\ and\ \citenamefont {Murakami}}]{Yoda2018-OME}%
  \BibitemOpen
  \bibfield  {author} {\bibinfo {author} {\bibfnamefont {T.}~\bibnamefont
  {Yoda}}, \bibinfo {author} {\bibfnamefont {T.}~\bibnamefont {Yokoyama}},\
  and\ \bibinfo {author} {\bibfnamefont {S.}~\bibnamefont {Murakami}},\
  }\bibfield  {title} {\bibinfo {title} {Orbital edelstein effect as a
  condensed-matter analog of solenoids},\ }\href
  {https://doi.org/10.1021/acs.nanolett.7b04300} {\bibfield  {journal}
  {\bibinfo  {journal} {Nano Letters}\ }\textbf {\bibinfo {volume} {18}},\
  \bibinfo {pages} {916} (\bibinfo {year} {2018})}\BibitemShut {NoStop}%
\bibitem [{\citenamefont {He}\ \emph {et~al.}(2020)\citenamefont {He},
  \citenamefont {Goldhaber-Gordon},\ and\ \citenamefont {Law}}]{He2020-OME}%
  \BibitemOpen
  \bibfield  {author} {\bibinfo {author} {\bibfnamefont {W.-Y.}\ \bibnamefont
  {He}}, \bibinfo {author} {\bibfnamefont {D.}~\bibnamefont
  {Goldhaber-Gordon}},\ and\ \bibinfo {author} {\bibfnamefont {K.~T.}\
  \bibnamefont {Law}},\ }\bibfield  {title} {\bibinfo {title} {Giant orbital
  magnetoelectric effect and current-induced magnetization switching in twisted
  bilayer graphene},\ }\href {https://doi.org/10.1038/s41467-020-15473-9}
  {\bibfield  {journal} {\bibinfo  {journal} {Nature Communications}\ }\textbf
  {\bibinfo {volume} {11}},\ \bibinfo {pages} {1650} (\bibinfo {year}
  {2020})}\BibitemShut {NoStop}%
\bibitem [{\citenamefont {Johansson}\ \emph {et~al.}(2018)\citenamefont
  {Johansson}, \citenamefont {Henk},\ and\ \citenamefont
  {Mertig}}]{IngridMerting-PhysRevB.97.085417}%
  \BibitemOpen
  \bibfield  {author} {\bibinfo {author} {\bibfnamefont {A.}~\bibnamefont
  {Johansson}}, \bibinfo {author} {\bibfnamefont {J.}~\bibnamefont {Henk}},\
  and\ \bibinfo {author} {\bibfnamefont {I.}~\bibnamefont {Mertig}},\
  }\bibfield  {title} {\bibinfo {title} {Edelstein effect in weyl semimetals},\
  }\href {https://doi.org/10.1103/PhysRevB.97.085417} {\bibfield  {journal}
  {\bibinfo  {journal} {Phys. Rev. B}\ }\textbf {\bibinfo {volume} {97}},\
  \bibinfo {pages} {085417} (\bibinfo {year} {2018})}\BibitemShut {NoStop}%
\bibitem [{\citenamefont {Dresselhaus}\ \emph {et~al.}(2007)\citenamefont
  {Dresselhaus}, \citenamefont {Dresselhaus},\ and\ \citenamefont
  {Jorio}}]{dresselhaus2007group}%
  \BibitemOpen
  \bibfield  {author} {\bibinfo {author} {\bibfnamefont {M.}~\bibnamefont
  {Dresselhaus}}, \bibinfo {author} {\bibfnamefont {G.}~\bibnamefont
  {Dresselhaus}},\ and\ \bibinfo {author} {\bibfnamefont {A.}~\bibnamefont
  {Jorio}},\ }\href {https://books.google.com.br/books?id=sKaH8vrfmnQC} {\emph
  {\bibinfo {title} {Group Theory: Application to the Physics of Condensed
  Matter}}}\ (\bibinfo  {publisher} {Springer Berlin Heidelberg},\ \bibinfo
  {year} {2007})\BibitemShut {NoStop}%
\end{thebibliography}


\end{document}